\documentclass[aps,prb,twocolumn,showkeys,superscriptaddress]{revtex4-2}
\usepackage{graphicx}
\usepackage{bm}
\begin{document}
\title{Photon mediated energy, linear and angular momentum transport in fullerene and graphene systems beyond local equilibrium}

\author{Jian-Sheng Wang}
\affiliation{Department of Physics, National University of Singapore, Singapore 117551, Republic of Singapore\smallskip\looseness=-2}
\affiliation{Laboratoire Charles Coulomb (L2C), UMR 5221 CNRS-Université de Montpellier, F-34095 Montpellier, France}
\author{Mauro Antezza}
\affiliation{Laboratoire Charles Coulomb (L2C), UMR 5221 CNRS-Université de Montpellier, F-34095 Montpellier, France}
\affiliation{Institut Universitaire de France, 1 rue Descartes, F-75231 Paris, France\smallskip\looseness=-2}


\date{21 July 2023, revised again 23 February 2024}

\begin{abstract}
Based on a tight-binding model for the electron system, we investigate the transfer of energy, momentum, and 
angular momentum mediated by electromagnetic fields among buckminsterfullerene (C$_{60}$) and graphene nano-strips.   
Our nonequilibrium Green's function approach enables calculations away from local thermal equilibrium where
the fluctuation-dissipation theorem breaks down.    For example, the forces between C$_{60}$ and current-carrying
nano-strips are predicted.    It is found that the presence of current enhances the van der Waals attractive 
forces.   For two current-carrying graphene strips rotated at some angle, the fluctuational force and torque are 
much stronger at the nanoscale compared to that of the static Biot-Savart law. 
\end{abstract}


\maketitle

\section{Introduction}
Energy transfer can be carried out in three forms -- conduction, convection, and radiation \cite{menguc-2020}.   Radiation is special
in that we do not need a material medium for the transfer.   Energy can be transmitted in a vacuum.   From the past half a century of work, it has been
established that energy transfers are enhanced when objects are in the near field \cite{Hargreaves69,Domoto70,Polder71}.   This has been verified by many
experiments  \cite{Kittel05,shen09,Ottens11,Kim15,Cui17,Kloppstech17} and theoretical calculations \cite{Volokitin07,Basu09,Song15,Henkel17,Biehs21}.   Such near-field effects have also found many 
applications \cite{Marconot21}.  

A related transport phenomenon is the transfer of momentum.  This is the origin of the van der Waals or London attractive forces \cite{London37} at short distances and Casimir \cite{Casimir48,Lifshitz_1956,Klimchitskaya09,Mauro17} or Casimir-Polder forces \cite{Casimir_Polder_1948,Milonni93}
at a larger distance when the finite speed of light is taken into account.  An atom above a dielectric surface is a 
classic problem that has been investigated extensively \cite{Casimir_Polder_1948,henkel_radiation_2002,Mauro05}.
Progress has been made on subtle effects of the temperatures of the bodies \cite{Mauro06,Mauro-JPhysA-06,Mauro-exp-07,Mauro-PRA-11,Youssef-23}. 

So far, even for the global nonequilibrium situations, most of the theoretical developments have been based on the assumption of local thermal equilibrium \cite{Lifshitz_1956,Polder71}, where each object still satisfies the
fluctuation-dissipation theorem. 
Systems driven by the electric current can be modeled by a Doppler shift of the equilibrium conductivities 
phenomenologically \cite{Duppen16,Morgado17,Shapiro17,Svintsov19}.  
The effect of the temperature gradient of the objects has only been investigated recently 
\cite{Messina16,Messina-2020,Reina-2023}.     These investigations couple
the heat radiation with the diffusion equation or Boltzmann transport theory, still at a macroscopic or mesoscopic level.  
Another approach to nonequilibrium transport is to modify the Bose function with chemical potential bias \cite{Callahan21}.
Our motivation here is to work at the microscopic level, starting with a model of matter as electrons hopping on some
(lattice) sites.    Thus, the nonequilibrium aspect can be handled from first principles, using the Keldysh nonequilibrium 
Green's function (NEGF) formalism \cite{Keldysh65,Haug08,Wang-review-1,Wang-review-2}.    
The drawback, of course, is computational complexity, which limits the method to small
systems at the nanoscale.    Along this line, 
a general photon transport theory for energy, momentum, and angular momentum has been developed under the
framework of the nonequilibrium Green's function formulation \cite{zhang_angular_2020,zhang_microscopic_2022,wang_transport_2023}.
It reduces to the usual fluctuational electrodynamics \cite{Krueger11prl,Krueger12prb,Bimonte17} 
if a local thermal equilibrium is valid.  The objects can be put out of local thermal
equilibrium by connecting to two or more baths at different
temperatures or chemical potentials, causing the objects to have heat
or electric current.   An out-of-equilibrium system breaks reciprocity even though the Hamiltonian is
still reciprocal in the sense $H^T = H$; here, $H$ is the single particle Hamiltonian matrix, 
and the superscript $T$ denotes matrix transpose. 

\begin{figure} 
  \centering
  \includegraphics[width=1.0\columnwidth]{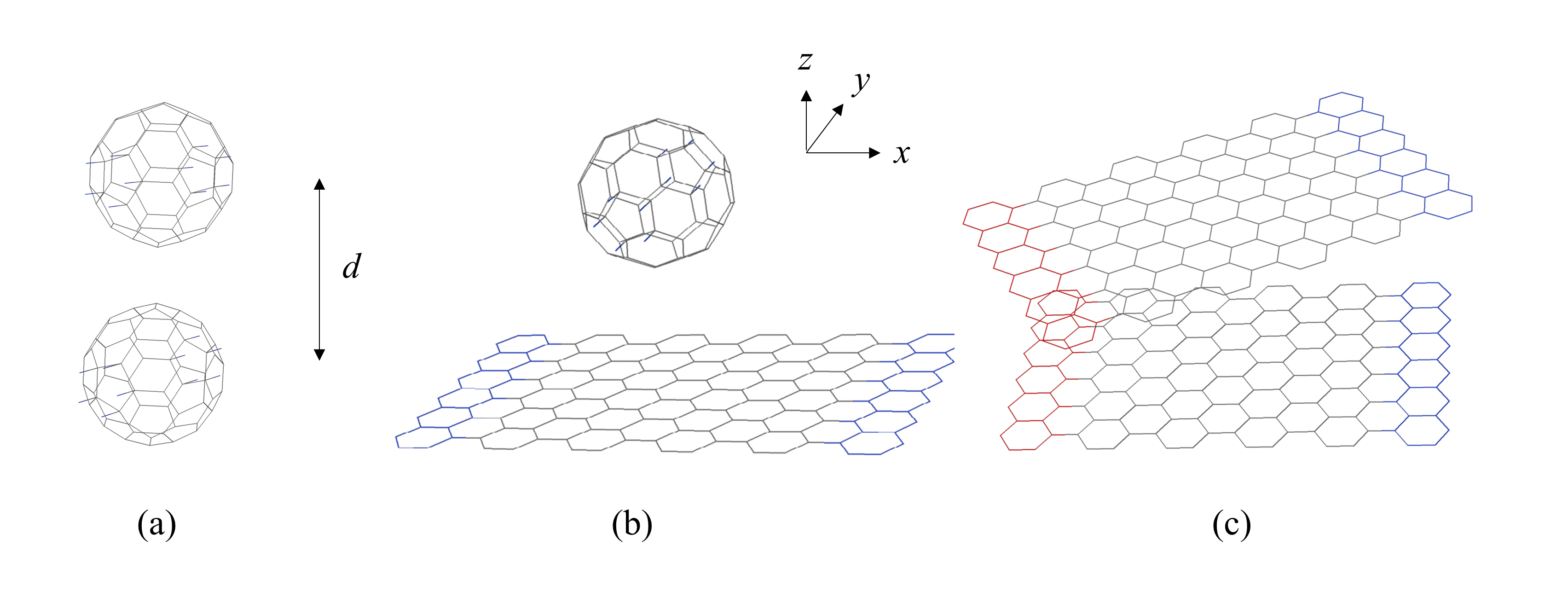}
  \caption{Schematic configurations of carbon systems. (a) The configuration of a C$_{60}$ molecule with 
  a center-to-center distance $d=1$\,nm away from another C$_{60}$. (b)  C$_{60}$ above a $13 \times 8$ armchair graphene strip.  The strip is approximately a square of length 1.5\,nm.  The C$_{60}$ is positioned at the center of 
the square.  (c) Two strips rotated by 30$^\circ$.   Only parts of the bath sites and bonds in color are shown 
explicitly.}
 \label{config}
\end{figure} 

In this paper, we demonstrate the power of the NEGF formalism for fullerene systems consisting of 
the C$_{60}$ molecules and graphene strips in the armchair geometry.  Specifically, we consider the transfer
of conserved quantities between C$_{60}$ and C$_{60}$ molecules,  C$_{60}$ with a flat graphene
strip, and between two strips rotated at an angle (see Fig.~\ref{config}).   

In the following, we give a recipe for calculation.  The derivations of these formulas are already presented in 
Refs.~\cite{zhang_angular_2020,zhang_microscopic_2022,wang_transport_2023}, except for more computational details.
We define the model considered, present the numerical results, and discuss their significance.  In the Appendices, we give a quick derivation of the Meir-Wingreen formula and relate it to 
the Casimir-Polder formula for van der Waals forces. 

\section{Model}
We consider systems as objects described by a tight-binding model of the form $\sum_{jk} c_j^\dagger H_{jk}^\alpha c_k$
of spinless free electrons.   Each object $\alpha$ is connected to a number of baths or electron reservoirs so that
the number of electrons on the objects can fluctuate, and the reservoirs define their thermal properties.  The
baths can also have chemical potential bias so that the object may have an electric current.   All this is described
by additional electron self-energies $\Sigma$ due to the baths so that the electron Green's functions for the
object $\alpha$ are \cite{Datta95}
\begin{eqnarray}
G^r_\alpha(E) &=& \left( E + i\eta - H^\alpha - \sum \Sigma^r(E) \right)^{-1},\\
G^{<,>}_\alpha(E) &=& G^r_\alpha(E) \sum \Sigma^{<,>}(E) G^a_\alpha(E),
\end{eqnarray}
where the first line is the retarded Green's function, and the second line is the lesser ($<$) and greater ($>$) Green's
functions which are given by the Keldysh equations.  The advanced Green's function is obtained by the Hermitian conjugate
of the retarded one,  $G^a = (G^r)^\dagger$, and the sum indicates the possibility of many baths for each object.
We assume that each bath is in thermal equilibrium so that we have the fluctuation-dissipation theorems for $\Sigma^{<,>}$.
That is, $\Sigma^< = -f (\Sigma^r - \Sigma^a)$, $\Sigma^> = (1-f) (\Sigma^r - \Sigma^a)$, where 
$f = 1/\bigl( \exp((E-\mu)/(k_BT)) + 1\bigr)$ is the Fermi function. 
Thus, each bath is characterized by two parameters: the temperature $T$ and the chemical potential $\mu$. 
In particular, to be consistent, the damping of the center, $\eta$, is also counted as one of the baths.
This central bath can be interpreted as the substrate or gate applied to the system. 

The electron-field interaction is described in the scalar potential $\phi=0$ gauge with the Peierls substitution 
Hamiltonian \cite{Peierls33}, $\sum_{jk} c^\dagger_j H_{jk} c_k \exp\left(-i \frac{e}{\hbar} \int_{{\bf R}_k}^{{\bf R}_j} {\bf A} \cdot d{\bf r} \right)$.  Here $c^\dagger_j$ is the creation operator of the electron at site $j$, $c_k$ is the annihilation
operator at site $k$, and $\bf A$ is the vector potential. 
The electron-photon interaction matrices can be obtained from the Hamiltonian and the locations of the sites,
via the velocity matrix:
\begin{equation}
{\bf V}_{jk}^\alpha = \frac{1}{i\hbar} H_{jk}^\alpha \left({\bf R}_j - {\bf R}_k\right),
\end{equation}
where ${\bf R}_j$ is the three-dimensional vector of the location of site $j$.    We introduce an ${\bf M}^l$ matrix so
that a sum over site $l$ produces $e {\bf V}^\alpha$, by 
\begin{equation}
\label{eqMmatrix}
M_{jk}^{l\mu} = \frac{1}{2} e \left( \delta_{lj} + \delta_{lk} \right) V_{jk}^{\alpha,\mu},\quad \mu = x,y,z,
\end{equation}
i.e., ${\bf M}^l$ is half of the $l$-th row and $l$-th column of the ${\bf V}^\alpha$ matrix times $e$, here $e$ ($>\!0$) is the magnitude of the 
elementary charge. 

\section{Photon self-energies and Meir-Wingreen formulas}
In the usual approach to fluctuational electrodynamics \cite{Rytov53,Rytov89}, the material property for electromagnetic response
is given by a frequency-dependent local dielectric function.   Such a description would not be suitable when
the system is at the nanoscale, such as a C$_{60}$ molecule.   In the NEGF approach, the dielectric 
function is replaced by nonlocal quantities (matrices in site $l$ and direction $\mu$) that we call self-energies.
These are the self-energies for the photon interacting with the electrons.  It is    
a key step in the calculation.  Under the random-phase approximation, they are given by the current-current 
correlations \cite{wang_transport_2023},
\begin{eqnarray}
\Pi^<_{l\mu,l'\nu}(\omega) = \qquad\qquad\qquad\qquad\qquad\qquad\qquad\qquad\quad \nonumber \\
  - 2 i \int_{-\infty}^{+\infty}\!\! \frac{dE}{2\pi} {\rm Tr} 
\bigl[ M^{l\mu} G^<(E) M^{l'\nu} G^>(E - \hbar \omega)  \bigr], \\
\Pi^r_{l\mu,l'\nu}(\omega) = \qquad\qquad\qquad\qquad\qquad\qquad\qquad\qquad\quad \nonumber \\
  -2 i \int_{-\infty}^{+\infty}\!\! \frac{dE}{2\pi} {\rm Tr} 
\Bigl[ M^{l\mu} G^r(E+\hbar \omega ) M^{l'\nu} G^<(E)  + \nonumber \\
  M^{l\mu} G^<(E) M^{l'\nu} G^a(E - \hbar \omega) \Bigr] .
\end{eqnarray}
The greater version $\Pi^>$ is obtained by swapping lesser with greater for the electron Green's functions.  
The Keldysh version is defined as $\Pi^K = \Pi^< + \Pi^>$, and advanced version is $\Pi^a = (\Pi^r)^\dagger$.
Here, the dagger means taking complex conjugate and transposing in both the space and the direction index, $(l,\mu)$. 
The extra factor of 2 is for the spin degeneracy.  The retarded self-energy is related to the conductivity in
a continuum description by $\sigma = i\Pi^r/\omega$ \cite{Mahan00}.

The self-energies, also known as polarizabilities, are the critical inputs for the transport calculations.
The numerical accuracy of the energy integration needs to be maintained to high standards.  The broadening 
parameter $\eta$ controls the integration spacings.  We choose this spacing a few times smaller than $\eta$,
and look for convergence. 

\subsection{Diamagnetic term}
The diamagnetic term is needed to correctly describe the plasmon physics of the electrons.   For
the retarded $\Pi^r$, this term has no imaginary part; thus, it is nondissipative, describing the motion of the
electrons in response to the external field.  

If we expand the Peierls substitution term to second order in the vector field, using the trapezoidal rule for the 
line integral, the extra interaction responsible for the diamagnetic term is
\begin{eqnarray}
H' &=& - { e^2 \over 8\hbar^2} \sum_{jk\mu\nu} 
c^\dagger_j H_{jk} c_k (A_j^\mu + A_k^\mu) (R_j^\mu - R_k^\mu) \times \nonumber \\
&&\qquad\qquad (A_j^\nu +A_k^\nu) (R_j^\nu - R_k^\nu).
\end{eqnarray}
The contribution to the Dyson expansion of the contour ordered Green's function 
$D_{\alpha\beta}({\bf r}, \tau; {\bf r}', \tau') = \bigl\langle T_\tau A_\alpha({\bf r},\tau) A_\beta({\bf r}',\tau')\bigr\rangle/(i\hbar)$ due to this interaction is (at the lowest order)
\begin{eqnarray}
&&\left(\frac{1}{i\hbar}\right)^2 \left\langle T_\tau  A_\alpha({\bf r}, \tau) \int d\tau'' H'(\tau'') A_\beta({\bf r}',\tau') \right\rangle \nonumber  \\
&=& {e^2 \over 8\hbar^4} \sum_{jk\mu\nu} \int d\tau'' \Bigl\langle T_\tau
A_\alpha({\bf r}, \tau) \Bigl[ A_j^\mu(\tau'') A_j^\nu(\tau'')  + \nonumber \\  
&& A_j^\mu(\tau'') A_k^\nu(\tau'') +  A_k^\mu(\tau'') A_j^\nu(\tau'')  + A_k^\mu(\tau'') A_k^\nu(\tau'') \Bigr] 
\times \nonumber \\ 
 && A_\beta({\bf r}', \tau') \Big\rangle 
 H_{jk} \bigl\langle c_j^\dagger(\tau'') c_k(\tau'') \bigr\rangle \times \nonumber \\
&& \bigl(R_j^\mu - R_k^\mu\bigr)\bigl(R_j^\nu - R_k^\nu\bigr).
\end{eqnarray}
Here $\tau,\tau',\tau''$ are Keldysh contour times, and $T_\tau$ is the contour order super-operator. 
To the lowest order, we can separate the averages in photon space from that of electron space.  The
next step is to apply Wick's theorem for the $\bf A$ field and express the field correlation by the photon 
Green's function.   To identify the self-energy, we write the final expression in the form
$D \Pi D$, as a convolution in contour times and space (as discrete sums).   The extra diamagnetic
self-energy is
\begin{eqnarray}
\Pi^{\rm dia}_{l\mu,l'\nu}(\tau_1, \tau_2) = \frac{i e^2}{2\hbar} \delta(\tau_1, \tau_2) 
\sum_{jk}\Bigl[  \delta_{lj} \delta_{l'j} + \delta_{lj} \delta_{l'k} + \nonumber  \\
 \delta_{lk} \delta_{l'j} + \delta_{lk} \delta_{l'k} \Bigl] \bigl(R_j^\mu - R_k^\mu\bigr)\bigl(R_j^\nu - R_k^\nu\bigr)
 H_{jk} G_{kj}^<(0) .
\end{eqnarray}
Here $G^<(t)$ is the lesser Green's function of the electron in the time domain. 
A factor of 2 has been multiplied for the spin degeneracy. 
The final result is proportional to a delta function in contour time.  This means that there is no contribution to 
$\Pi^<$ and $\Pi^>$.  We only pick up a retarded component, $\Pi^r = \Pi^t - \Pi^<$, which is a constant in 
the frequency domain, due to the delta function $\delta(t_1 -t_2)$ in the time domain. 
Here $\Pi^t$ is the time-ordered version of the self-energy.   On a continuum, this
diamagnetic term is $-e^2 n/m$; here, $n$ is electron density.   It is also equal to the negative of 
current-current correlation at zero frequency, $-\Pi^r(\omega=0)$, from gauge invariance.  
On a lattice, it is somewhat more complicated due to our use of Peierls substitution Hamiltonian. 

\subsection{Free-field photon Green's function}
The photon Green's function $D = - \mu_0 G$ is the same as the usual dyadic Green's function $G$ up to a numerical 
factor of $-\mu_0$, which is the vacuum permeability.  In a vacuum without matter, we use the symbol $v$ to denote
the free photon Green's function, which is given explicitly by the formula
\cite{Novotny06,Keller11}
\begin{eqnarray}
\label{eq-v-real-space}
v^r({\bf r},\omega) &=& 
-\frac{e^{i \frac{\omega}{c} r}}{4\pi \epsilon_0 c^2 r} \Bigg\{ 
\bigl( \stackrel{\leftrightarrow}{\bf U} - \hat{\bf R}\hat{\bf R} \bigr) +  \nonumber \\
&& \biggl[ - \frac{1}{ i\frac{\omega}{c}r} + 
\frac{1}{\left( i \frac{\omega}{c}r \right)^2}\biggr]
\bigl( \stackrel{\leftrightarrow}{\bf U} - 3 \hat{\bf R}\hat{\bf R} \bigr) \Bigg\}. 
\end{eqnarray} 
Here $\stackrel{\leftrightarrow}{\bf U}$ is the identity dyadic,  $\hat{\bf R} = {\bf r}/r$ is the radial direction unit vector. 

To compute the force and torque, we also need the derivative of this expression with respect to space.
We obtain in component form a messy formula as
\begin{eqnarray}
\label{eq-dv} 
\partial_\mu v^r_{\alpha\beta} &=& \frac{e^X}{4\pi\epsilon_0 c^2 r^2} \Biggl[ 
\left( 2 - X - \frac{3}{X}+ \frac{3}{X^2}\right) \hat{R}_\mu \delta_{\alpha\beta} + \nonumber \\
&&\left( -6 + X + \frac{15}{X}- \frac{15}{X^2}\right)  \hat{R}_\mu \hat{R}_\alpha \hat{R}_\beta  + \nonumber \\
&&\left( 1 - \frac{3}{X} + \frac{3}{X^2}\right) \bigl( \delta_{\mu\alpha} \hat{R}_\beta + \delta_{\mu\beta} \hat{R}_\alpha\bigr)  \Biggr],
\end{eqnarray}
where $X = i \omega r/c$, and $\mu$, $\alpha$, or $\beta$ takes the $x$, $y$, or $z$ direction.  $\hat{R}_\mu$ is the 
component of the unit vector $\hat{\bf R}$ in the $\mu$ direction.

In computing the transported quantities with the Meir-Wingreen formulas, we only need the values of $D$ at 
the electron sites.  Hence, the free Green's function is also evaluated at a discrete set of distances 
${\bf r} =  {\bf R}_j - {\bf R}_k$, where ${\bf R}_j$ and ${\bf R}_k$ are the tight-binding sites.   The Dyson equation
$D^r = v^r + v^r \Pi^r D^r$ is a $3N \times 3N$ matrix equation for $N$ electron sites.   The extra factor of 3 is due to the 
$x$, $y$, and $z$ directions for each site.  The two formulas, Eq.~({\ref{eq-v-real-space}) and (\ref{eq-dv}), diverge 
at ${\bf r}=0$, and thus cannot be used.   The divergence is due to our approximation of electrons to be point-like.
In reality, the wave functions of the electrons are extended with sizes of order angstrom.  To deal with this divergence,
we need a cut-off distance $r_c = 2$ Hartree atomic units (a.u.).  When $r < r_c$, we reset $r$ to be $r_c$ with $\hat{\bf R} = 0$.   This gives, following Refs.~\cite{yaghjian_electric_1980,Abajo-2012},  
$v^r({\bf 0}, \omega) \approx \frac{1}{4\pi\epsilon_0 \omega^2 r_c^3}\stackrel{\leftrightarrow}{\bf U}$, and $\nabla v^r = 0$.   
The results are sensitive to the value $r_c$ as it reflects the screening in the dielectric matrix
$\epsilon = 1 - v^r \Pi^r$.   The value $r_c$ is determined from the Coulomb energy of two 
overlapping $p$-orbitals. With the choice of $r_c = 2\,$a.u., it gives a reasonable screening strength for carbon atoms
with a static dielectric constant close to 3 (as compared by the screened and bare polarizability $\alpha$ of C$_{60}$). 

\subsection{Dyson equation and Meir-Wingreen formulas}

With these preparations, we solve the Dyson equation,
\begin{equation}
D^r = v^r + v^r \sum_\alpha \Pi^r_\alpha D^r,
\end{equation} 
and calculate the derivatives by ${\bm \nabla} D^r = {\bm \nabla} v^r + {\bm \nabla} v^r \sum_{\alpha} \Pi^r_\alpha D^r$.   

We also need the Keldysh version for distribution, which is obtained by $D^K = D^r \sum_\alpha \Pi^K_\alpha D^a$, 
and the advanced version by $D^a = (D^r)^\dagger$.  In calculating the Keldysh equation, we
have ignored one term, the bath at infinity, $\Pi_\infty^K$.  This term represents the dissipation of energy to
infinity.   When several objects are close, the magnitudes of their energy transfer are much larger than those transferred to infinity, so omitting
it is justifiable.   We can have another formula for these emitted to infinity by integrating the Poynting vector or
Maxwell's stress tensor on a sphere.  We note that conservation laws break down in our approximation here.

Finally, the Meir-Wingreen formulas for the energy transferred out from object $\alpha$, force and torque applied
to object $\alpha$, are \cite{wang_transport_2023},
\begin{equation}
\label{eq-O-meir-wingreen}
{d \langle \hat{O}\rangle \over dt} = {\rm Re} \int_0^\infty \frac{d\omega}{2\pi} {\rm Tr}\Bigl[ 
\hat{O} \bigl( D^r \Pi^K_\alpha + D^K \Pi_\alpha^a\bigr) \Bigr],
\end{equation}
where $\hat{O}$ is the operator $-\hbar \omega$ for energy transfer, $-i\hbar {\bm \nabla}$ for force, 
and ${\bf r} \times (-i \hbar {\bm \nabla})+ {\bf S}$ for torque \cite{strekha22}.  Here $S_{\mu\nu}^{\gamma} = 
(-i\hbar) \epsilon_{\mu\nu\gamma}$ is the spin angular momentum operator expressed in the Levi-Civita symbol.
These operators act on the first argument of space and direction of the photon Green's function $D$. 
The trace is a sum of the discrete site indices and the directions. 

We comment that if the system has an overall thermal equilibrium in the sense $\Pi^K = (2N+1)(\Pi^r - \Pi^a)$ and
similarly for $D^K$, we can transform 
the factor $D^r \Pi^K_\alpha + D^K \Pi^a_\alpha$ to $2 i (2N+1) {\rm Im} (D^r \Pi_\alpha^r)$ (for reciprocal systems).  
Here $N = 1/\bigl(\exp(\hbar \omega/(k_B T)) -1\bigr)$ is the Bose function. This will give
0 energy currents, but the force and torque are not zero.   We can further transform ${\rm Im} (D^r \Pi_\alpha^r)$
into the Matsubara frequencies, $\omega \to i\omega_n$, making a contact with the Liftshitz theory for 
Casimir force \cite{Lifshitz_1956,wylie_quantum_1984,henkel_radiation_2002,Bimonte22}. 
The formulas should be numerically more stable in the imaginary frequencies as the functions are not oscillatory.
However, the drawback is that we can no longer handle nonequilibrium problems.  The Matsubara sum form for the 
equilibrium system is then
\begin{equation}
{d \langle \hat{O}\rangle \over dt} =  \frac{2k_B T}{\hbar} \sum'_{n\ge 0} {\rm Tr}\Bigl[ 
(i \hat{O} )  D^r(i\omega_n) \Pi^r_\alpha(i\omega_n) \Bigr],
\end{equation}
where the prime means that the $n=0$ term has a weight of $1/2$.  The Matsubara frequencies take the values $\omega_n = 2\pi n k_B T/\hbar$. 

Since the derivative of $v^r$ is antisymmetric, any symmetric matrix multiplying it and taking trace is 0.  Using this property,
for reciprocal systems, we can omit the ${\bm \nabla} v^r \Pi^r_\alpha + {{\bm \nabla} v}^r \Pi^r_\alpha D^r \Pi^r_\alpha$ terms, and use only 
${\bm \nabla} v^r \sum_{\beta \neq \alpha} \Pi^r_\beta D^r \Pi^r_\alpha$ \cite{Gomez-Santos09}.   The omitted terms, which constitute the self-interaction force, are zero for equilibrium systems.
This improves numerical stability for the force calculation.

\section{C$_{60}$--C$_{60}$ van der Waals force}

\begin{figure}
  \centering
  \includegraphics[width=0.8\columnwidth]{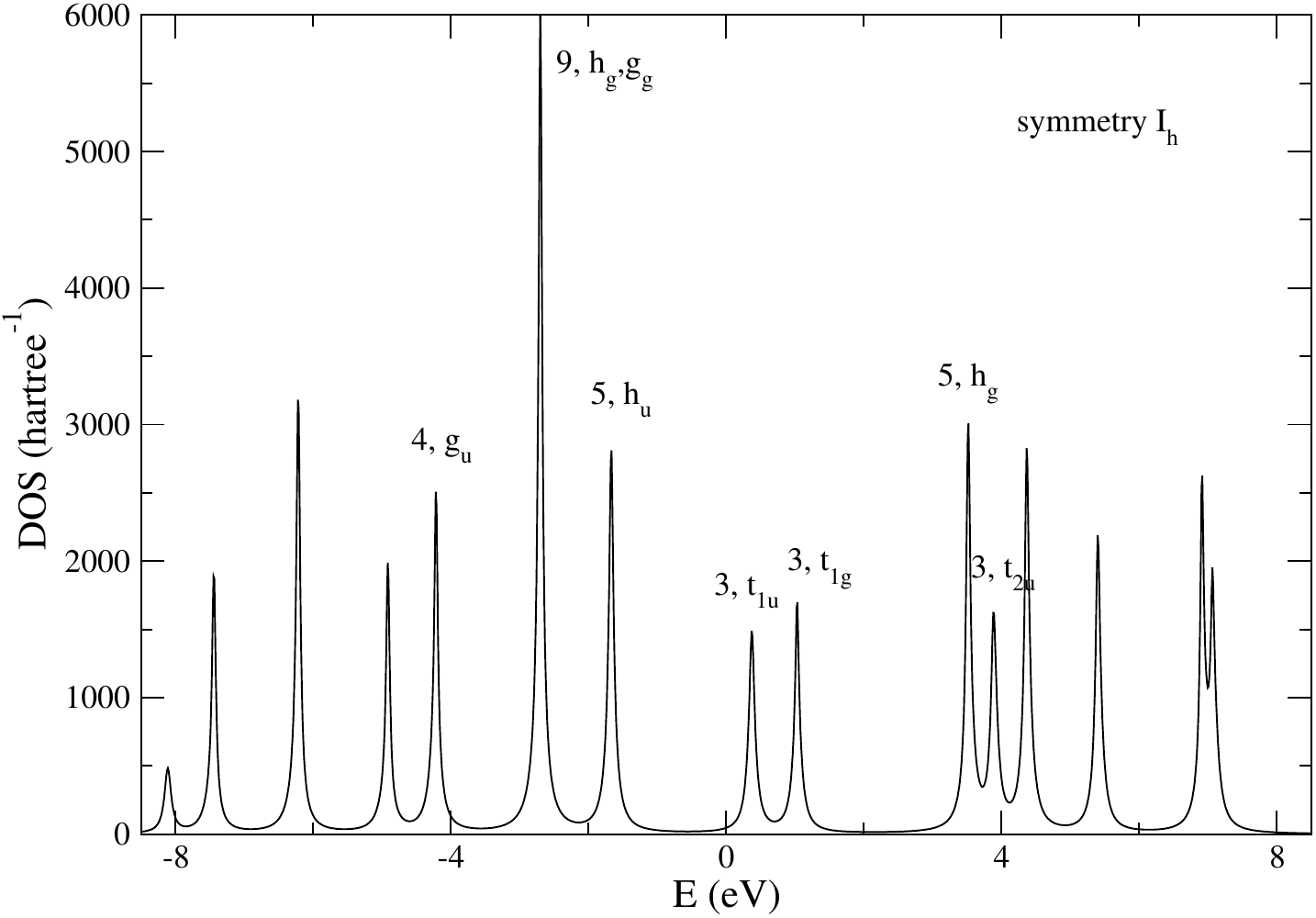}
  \caption{The density of states of C$_{60}$.  Some of the peaks are marked by the degeneracy and the orbital symmetry labels.}
 \label{C60-DOS}
\end{figure} 

\begin{figure}
  \centering
  \includegraphics[width=0.8\columnwidth]{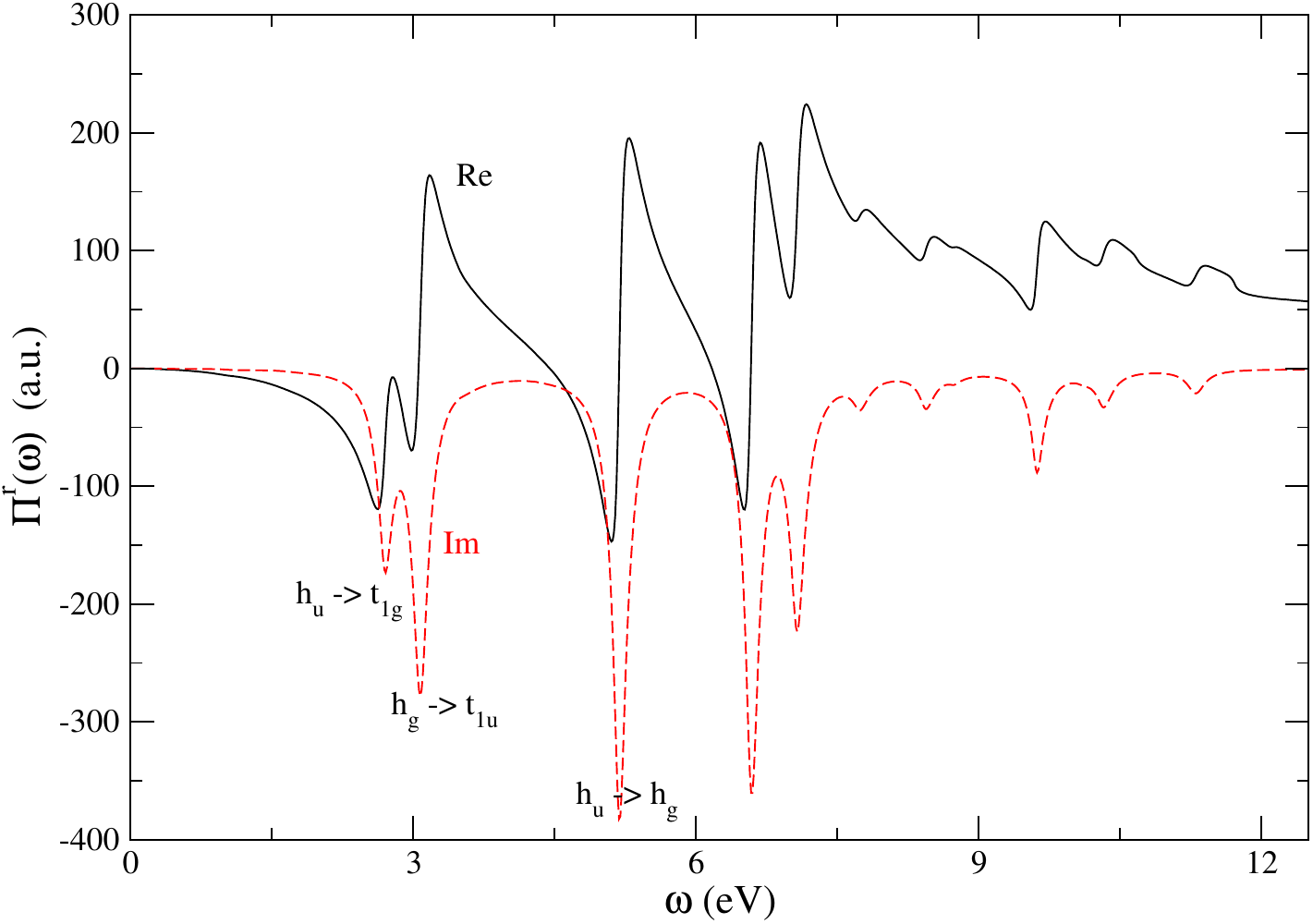}
  \caption{Sum total of the retarded photon self-energy $\Pi^r(\omega)$ in atomic units.  The solid line is for the real part, and the dashed line is 
  for the imaginary part. Some of the peaks are marked with corresponding electron state transitions.}
 \label{C60-Pi_r}
\end{figure} 

We use the coordinates for C$_{60}$ of Ref.~\cite{senn_computation_1995}, assuming a nearest neighbor hopping of $t=2.7\,$eV \cite{satpathy-1986}.   The molecule is rotated 90$^\circ$ along the $x$ direction 
so that the $z$ axis becomes $y$ in our coordinate system, see Fig.~\ref{config}.

The density of states (DOS) and the calculated $\Pi^r$ are presented in Fig.~\ref{C60-DOS} and \ref{C60-Pi_r}.
The C$_{60}$ molecule has the symmetry group I$_h$; thus, the states are highly degenerate.  Some states 
near the Fermi energy $\mu = 0\,$eV are labeled by their degeneracy and orbital symmetry.   The lowest occupied
and highest unoccupied states have a gap of $2.04\,$eV.  The DOS is calculated
according to $- {\rm Im}\, {\rm Tr}\, G^r(E)$.   A damping of $\eta = 27\,$meV in the electron Green's function 
is used throughout all the calculations in this paper. 
The corresponding self-energy $\Pi^r_{\rm tot}(\omega) = \sum_{j,k,\mu} \Pi_{j\mu,k\mu}^r(\omega)$ is 
the sum total, i.e., summed over the sites and traced over the direction.  We expect $\Pi^r(\omega) \sim \omega^2$
for small frequency, and $3Ne^2/m$ at $\omega \to \infty$.  Here $e$ is the elementary charge unit, $m$ is the effective mass of an electron, and $N$ is the number of electrons. 
The retarded susceptibility $\chi^r= \sum \Pi^r (1 - v \Pi^r)^{-1}$ is related to the molecule's dynamic polarizability, $\bar{\alpha}(\omega)
= -\frac{1}{3} \chi^r(\omega)/\omega^2$.   The induced dipole moment is related to the applied electric field by
${\bf p} = \bar{\alpha}(\omega) {\bf E}$, which defines $\bar{\alpha}(\omega)$.   The static polarizability of 
a C$_{60}$ molecule computed by fitting the $\omega^2$ dependence of $\chi^r$ for small $\omega$ gives 
$\bar{\alpha}(0) \approx 460\,$a.u.   The value is consistent with first principles and experimental results
\cite{wang_hyperpolarizability_1993,zope_static_2008,KU-Lao21}. 

As a check, we consider two identical C$_{60}$ molecules with a center-to-center distance $d$ in thermal equilibrium.  Ten baths are 
weakly coupled to the C$_{60}$ molecules with two of the opposite side pentagons, each connected to
independent one-dimensional chains
with a coupling strength $\Gamma = -2 {\rm Im}\, \Sigma^r =  0.4\,$eV to serve as baths.  The result of the total van der Waals force between the two 
C$_{60}$ molecules is plotted
in Fig.~\ref{C60-force}.   The result is calculated at 300\,K.   We compared with the result from the sum of the Matsubara frequency 
method valid for overall equilibrium systems.   It is seen that the Matsubara method is more accurate for large distances. 
We also checked the temperature dependence. As expected, the force is insensitive to temperature.  From 30\,K to 1000\,K, the force
changes by only about 3\%.

\begin{figure}
  \centering
  \includegraphics[width=0.8\columnwidth]{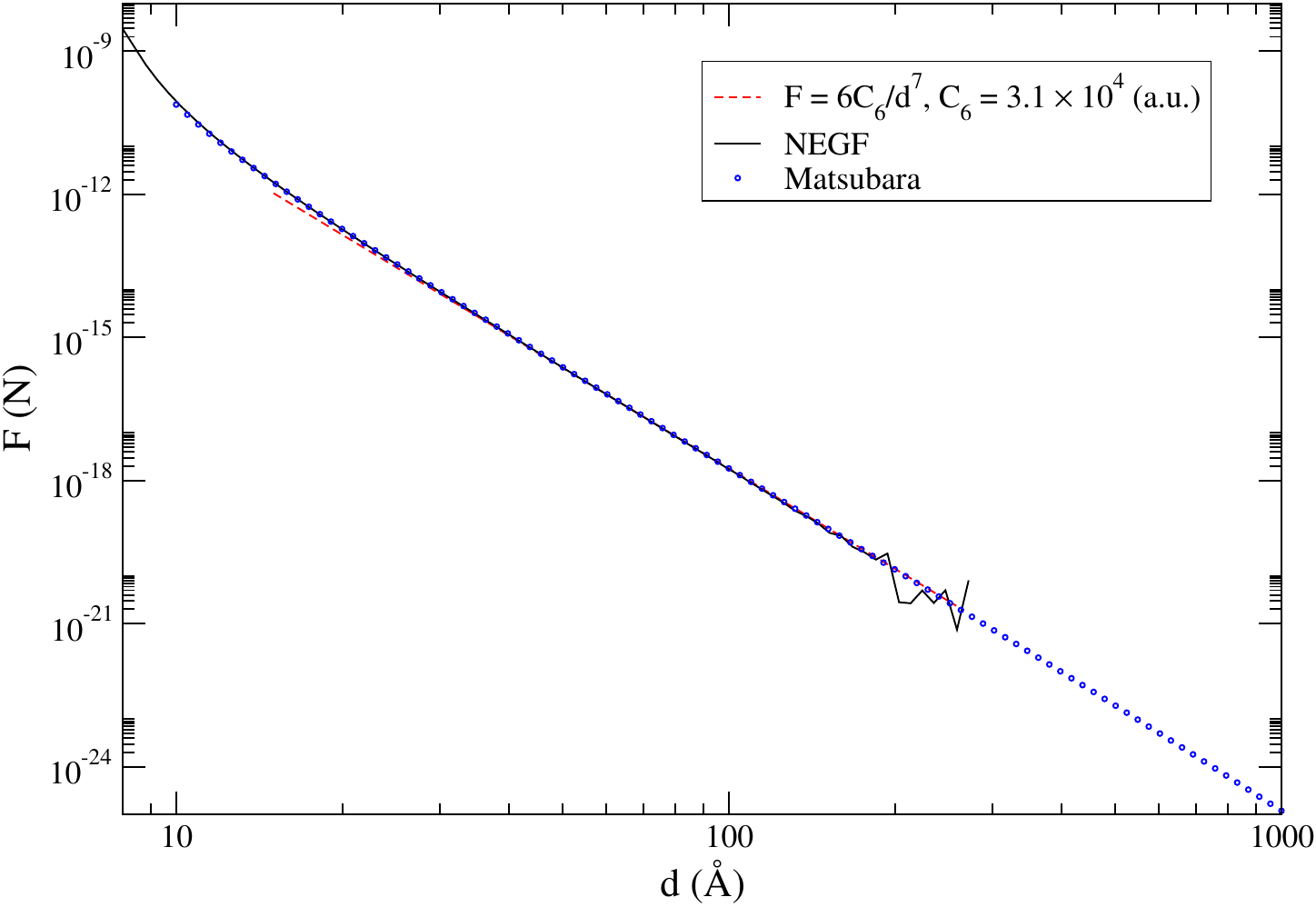}
  \caption{The force between two C$_{60}$ molecules with center to center distance $d$ at 300\,K.   The solid line is by the NEGF method; 
  circles are from the Matsubara method.  The dashed line is the asymptotic 
  force of $6 C_6/d^7$ with a fitted value of the coefficient $C_6$ for a van der Waals potential.}
 \label{C60-force}
\end{figure} 

The value of the van der Waals coefficient $C_6$ is found to be $31.3$ k\,hartree (bohr)$^6$  (i.e., a.u.).  This result is about $1/3$ of the first principle calculations \cite{KU-Lao21,Kauczor13}.   The smallness is due to our model.   In our
tight-binding model, we have one electron per carbon for the $\pi$ orbitals, and the $\sigma$ orbitals are absent.   Therefore, the transitions
of these orbitals to delocalized higher energy states are missing in our treatment.   Thus, our force calculations
are only qualitative.  
It appears that the full dynamic polarizability is important to determine the van der Waals force quantitatively.  We have also considered other orbitals still within the tight-binding models, the $\sigma$ orbitals as well as unoccupied $d$-orbitals \cite{Panhui-vdW}. These are important but very expensive to calculate, in order to bring our result to be in agreement with density functional theory-based calculations. Thus, the $\pi$-orbital-only model does have a fundamental limitation as far as van der Waals force is concerned.

\section{Energy and momentum transfer between C$_{60}$ and graphene strip}

The graphene nanoribbon is modeled with a nearest neighbor hopping model with the same hopping
parameter $t = 2.7\,$eV as for C$_{60}$ on a part of a honeycomb lattice, with a bond length $a=1.4\,$\AA.   Conceptually, we use 
an infinitely long strip with an armchair edge.    We cut a section of it (see Fig.~\ref{config}) as exposed and have electromagnetic
interactions, while the remaining two semi-infinite segments serve as baths.  The reason for using armchair edges 
instead of zigzag edges, due to the existence of localized edge states in zigzag configuration, is numerical stability.  The qualitative features 
are the same as zigzag edges.       

\begin{figure}
  \centering
  \includegraphics[width=0.8\columnwidth]{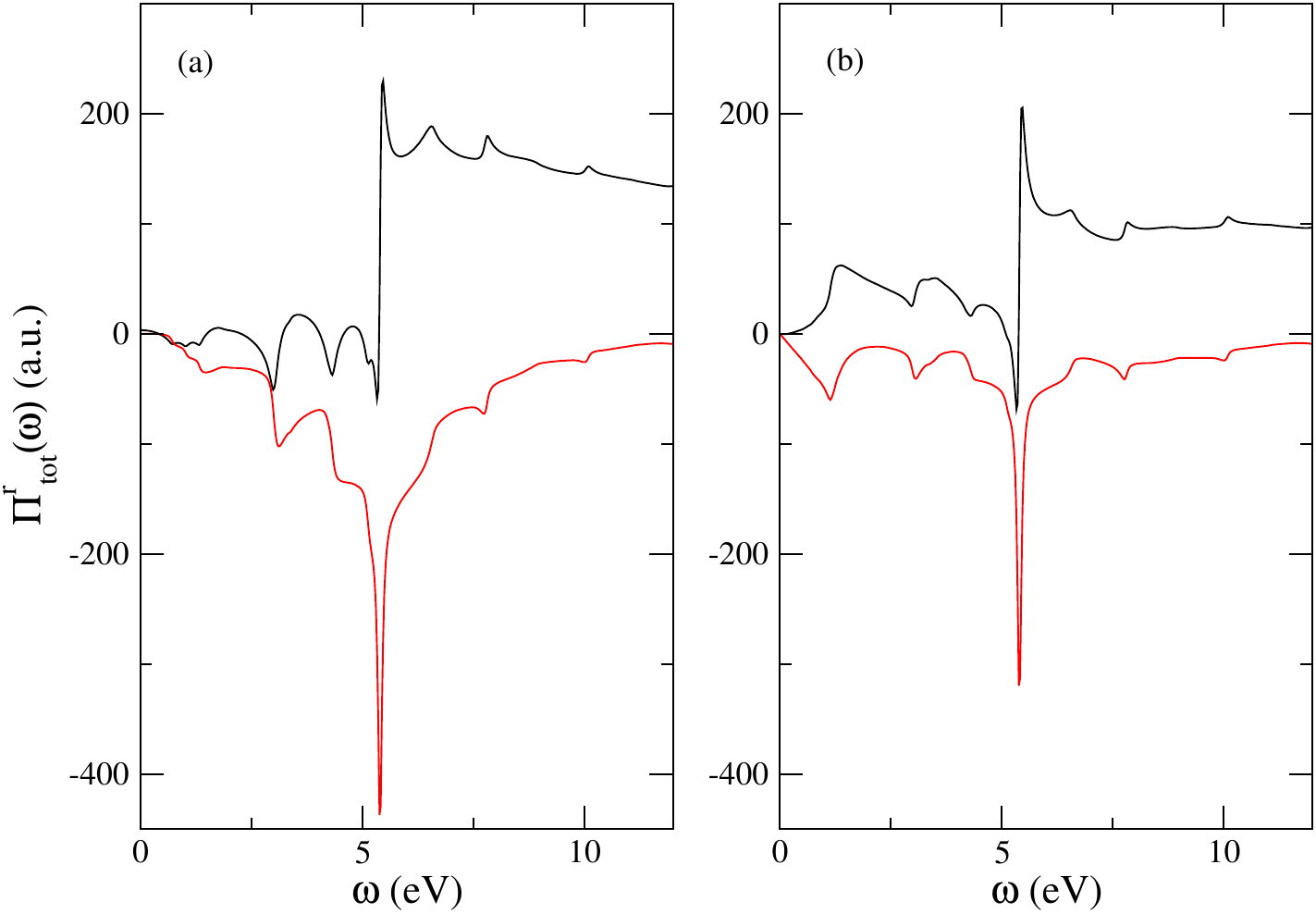}
  \caption{Sum total of the retarded photon self-energy $\Pi^r(\omega)$ in atomic units for the $13 \times 8$ 
armchair strip with baths.  The black line is 
for the real part, and the red line is for the imaginary part. (a) At chemical potential $\mu = 0\,$eV, (b) 
$\mu=4\,$eV.}
 \label{Pi_r13x8}
\end{figure} 

Figure~\ref{Pi_r13x8} is an analogous plot as Fig.~\ref{C60-Pi_r} for the $13 \times 8$ nanostrip.   Compared 
to the C$_{60}$ self-energy $\Pi^r$, it is smooth instead of several sharp peaks.   This is because the spectrum
of the electron of an infinitely long strip (due to the two semi-infinite baths) has a continuum.  The high peaks
are caused by transitions between the energy $-t$ and $+t$ states.   At low frequencies at $\mu = 0\,$eV (undoped graphene)
the values of $\Pi^r$ are small, while at high chemical potential, it is in a metallic regime, and the value is relatively high.

When each object is in local equilibrium, i.e., the fluctuation-dissipation theorems are valid, the NEGF formulation
is equivalent to fluctuational electrodynamics.   This means that, in our model, each object is connected to
baths that are at the same temperatures and chemical potentials.  Then we have 
$\Pi^K_\alpha = (2 N_\alpha + 1) (\Pi^r_\alpha - \Pi^a_\alpha)$.  Together with reciprocity, 
$H^T = H$, which implies $(G^{r,a,<,>})^T = G^{r,a,<,>}$, and $(D^<)^T = D^<$, and
$D^a = (D^r)^*$, we can write the force as a sum of an overall equilibrium contribution and
a correction due to the nonequilibrium temperature effect, as 
\begin{eqnarray}
\label{eqforce} 
{\bf F}_\alpha = \int_0^\infty \frac{d\omega}{\pi} {\rm Tr\,} \Bigl[
\hbar\, {\rm Im} \bigl({\bm \nabla} D^r \Pi^r_\alpha\bigr) (2 N_\alpha + 1) 
\Bigr] +   \quad\qquad \\
 \qquad\int_0^\infty \frac{d\omega}{\pi} {\rm Tr\,} \left[
\hbar\, {\rm Im }\Bigl( \sum_{\beta \neq \alpha} \delta N_\beta
{\bm \nabla} D^r (\Pi_\beta^r - \Pi_\beta^a) D^a \Pi_\alpha^a \Bigr) 
\right], \nonumber
\end{eqnarray}
where $\delta N_\beta = N_\beta - N_\alpha$.   At short distances, it is dominated by the first term
as $\delta N_\beta$ is small comparing to $2N_\alpha + 1$. 

\begin{figure} 
  \centering
  \includegraphics[width=0.9\columnwidth]{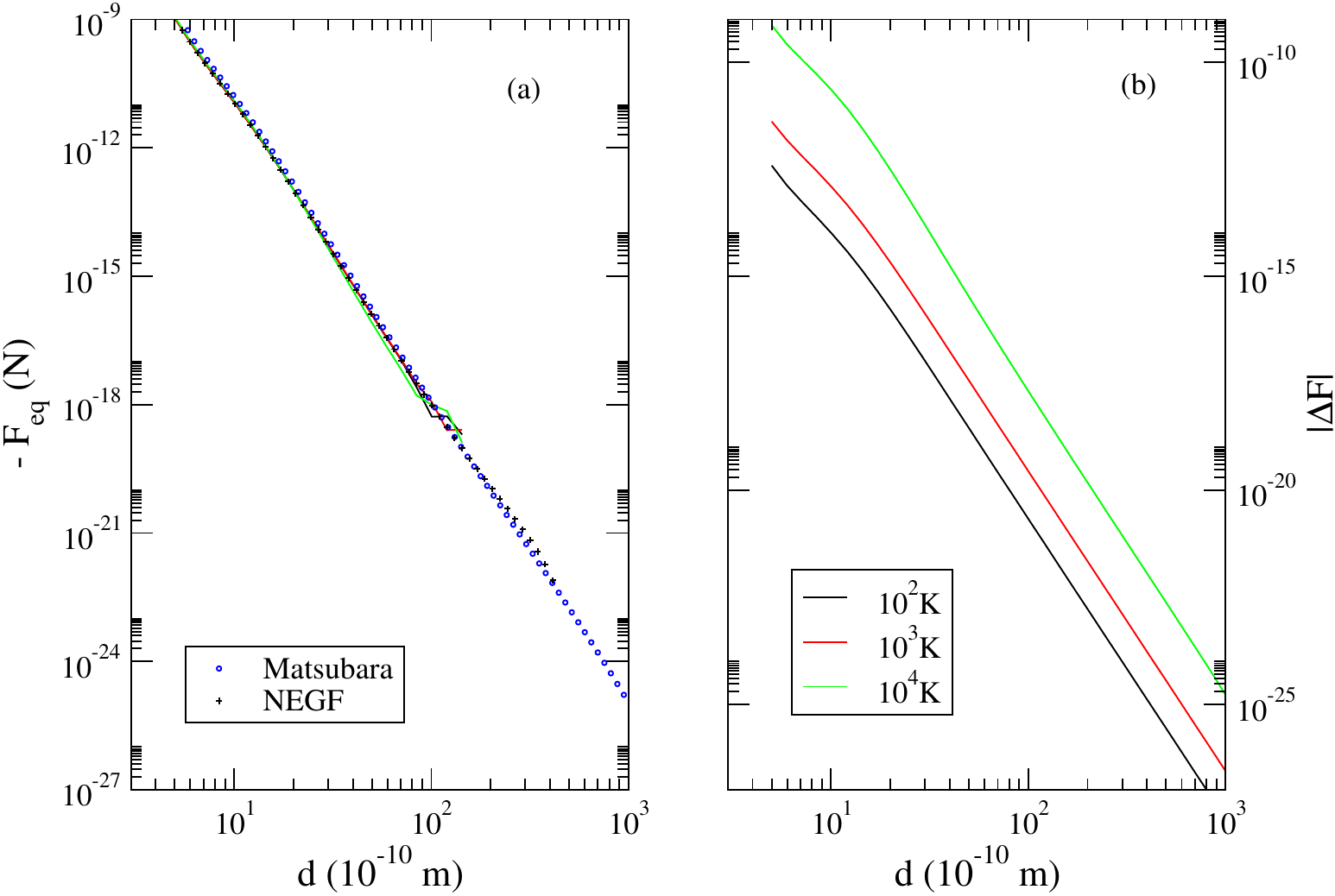}
  \caption{Force acting on C$_{60}$ by graphene.  (a) Force contribution from the first equilibrium term of
Eq.~(\ref{eqforce}).  Pluses by NEGF and circles from Matsubara method at $300\,$K.   The overlap curves follow the color scheme in (b) for the temperatures of the strip.  (b) The nonequilibrium contribution
from the second term.  The line color indicates strip temperatures.}
 \label{strip-C60-force}
\end{figure} 

In Fig.~\ref{strip-C60-force}, we plot the separate contributions from the first term and second term in the force formula
above.  The temperature of C$_{60}$ is set at 300\,K, while the strip is at 100\,K (black), 1000\,K (red),
and $10^4\,$K (green).  The left figure (a) is the equilibrium term.  
Although we call it equilibrium, strictly speaking, this is
not so as the temperatures of the strips are different from that of C$_{60}$, but the results are insensitive to 
temperature due to the $1$ in $2N_\alpha + 1$.  The self-energies of the two objects are only weakly dependent on the temperatures.
There is no correction term if the strip is also at 300\,K.  The correction terms are generally quite small unless the temperature is comparable to the eV energy scale.   The sign of
the correction is practically determined by $\delta N_\beta$;  it has the same sign as that of the first term if 
the strip is at a higher temperature than C$_{60}$, and opposite to that of the first term when the strip is at a lower 
temperature.  Both terms decay with distance as $d^{-7}$. 
 We could not see the effect of ``optical pressure'' at these distances. 

In a nonequilibrium setting where the local fluctuation-dissipation theorem breaks down, even when the Hamiltonian
is reciprocal, $H^T = H$, the Green's functions $G^{>,<}$ and self energies $\Pi^r$ are no longer reciprocal.  This
makes a numerical evaluation of the transport quantities rather unstable.   There is a large cancellation effect
between the first term $D^r \Pi_\alpha^K$ and the second term $D^K \Pi_\alpha^a$ in Eq.~(\ref{eq-O-meir-wingreen}); they are not simply related to
cancel some of the terms, i.e., a Caroli form of the type $\Pi_\beta \cdots \Pi_\alpha$ does not exist.   We make sure, numerically to high precision, the known identities are
satisfied, such as
$G^r - G^a = G^> - G^<$, $\Pi^r - \Pi^a = \Pi^> - \Pi^<$, and the optical theorem 
$D^r - D^a = D^r (\Pi^r - \Pi^a) D^a$.

\begin{figure} 
  \centering
  \includegraphics[width=0.9\columnwidth]{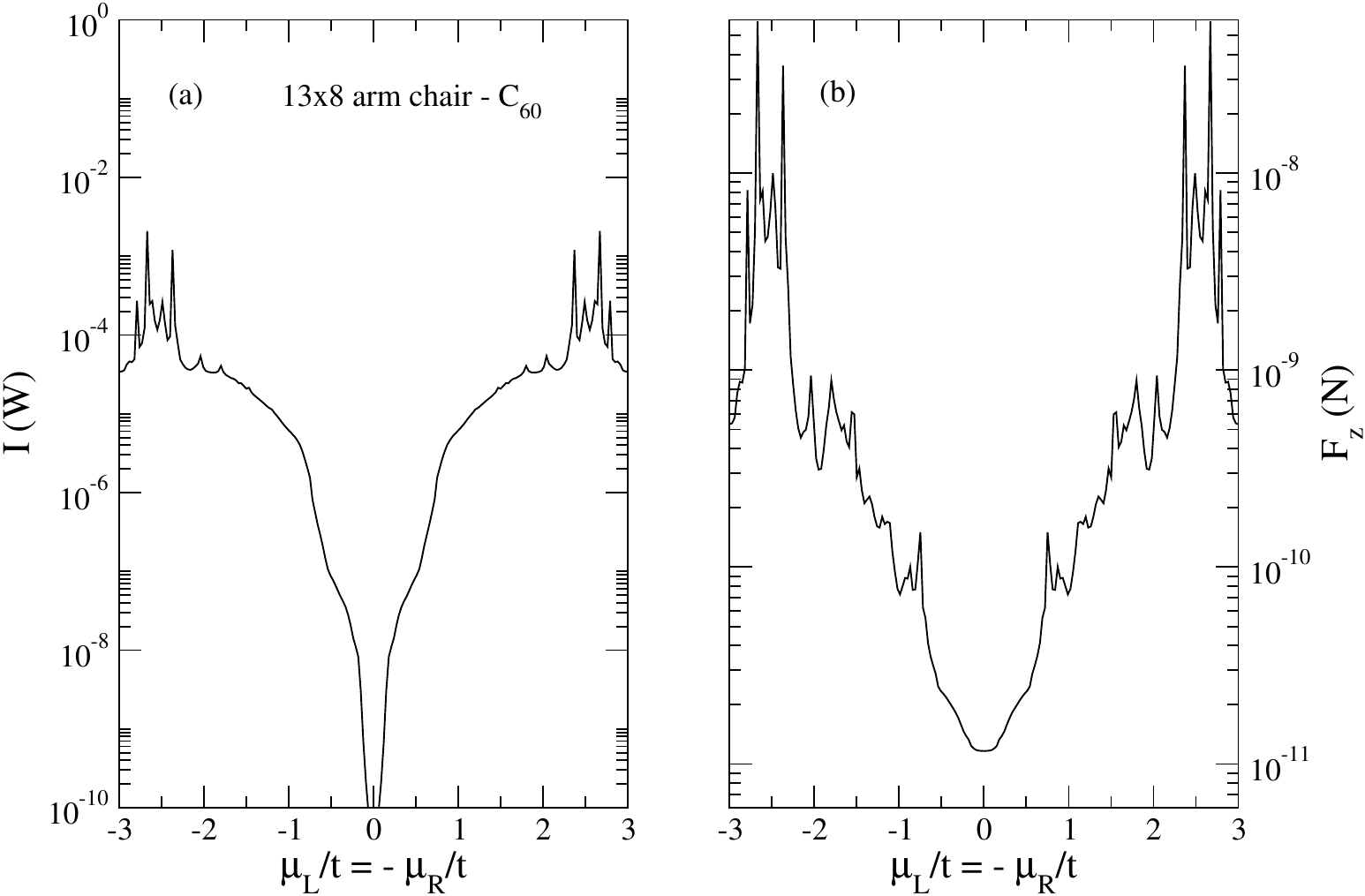}
  \caption{Transport between $13 \times 8$ graphene
strip of armchair edge with C$_{60}$ molecule with a center distance $d=1\,$nm away from the graphene at 300\,K.  
The strip is symmetrically biased ($\mu_L = -\mu_R$).   (a) Energy transferred to $C_{60}$,  and (b)  force in $z$ direction to graphene. 
}
 \label{strip-C60-sym}
\end{figure} 

The results presented above in Fig.~\ref{strip-C60-force} are for the cases where chemical potentials are kept at a constant
$\mu = 0\,$eV for all the baths.
In Fig.~\ref{strip-C60-sym}, we show the situation when the graphene experiences a 
symmetric bias ($\mu_L = - \mu_R$) of the chemical potentials from the two ends of the graphene strip.
The energy transferred to C$_{60}$ and the force applied to the strip are plotted against
the chemical potential of the left bath $\mu_L$ normalized by the hopping parameter $t$; the right bath is set to have the equal
and opposite value.   Although the temperatures of the two objects are at the same 300\,K,
the chemical potential bias causes 
the transfer of energy, which is a nonequilibrium effect.   The current flow generates radiation which is transferred to the molecule.   The molecule itself is in local thermal equilibrium.  The force is of the van der Waals repulsive 
type.   Newton's third law of the force is valid at our level of approximation.   The force
applied to C$_{60}$ is exactly equal and opposite to that of graphene plotted to high
precision.   The value is the smallest (in fact, 0 for the energy transfer) when there is no
bias, and it is symmetric about the bias and increases with high biases.   At huge 
biases, we see sharp peaks.  These are caused by some resonance of the adsorption/emission. 
Granted, these values of bias are unrealistic as the system will melt before that.
The force in $y$ direction in the graphene plane, perpendicular to the
current, is very small ($10^{-15}\,$N) due to the structural symmetry.   The force in $x$
direction, the direction of the driven current, is also small (peaked at about $10^{-13}\,$N), 
but it is an odd function of
$\mu_L$, reflecting the directionality of the current.   

\begin{figure} 
  \centering
  \includegraphics[width=0.9\columnwidth]{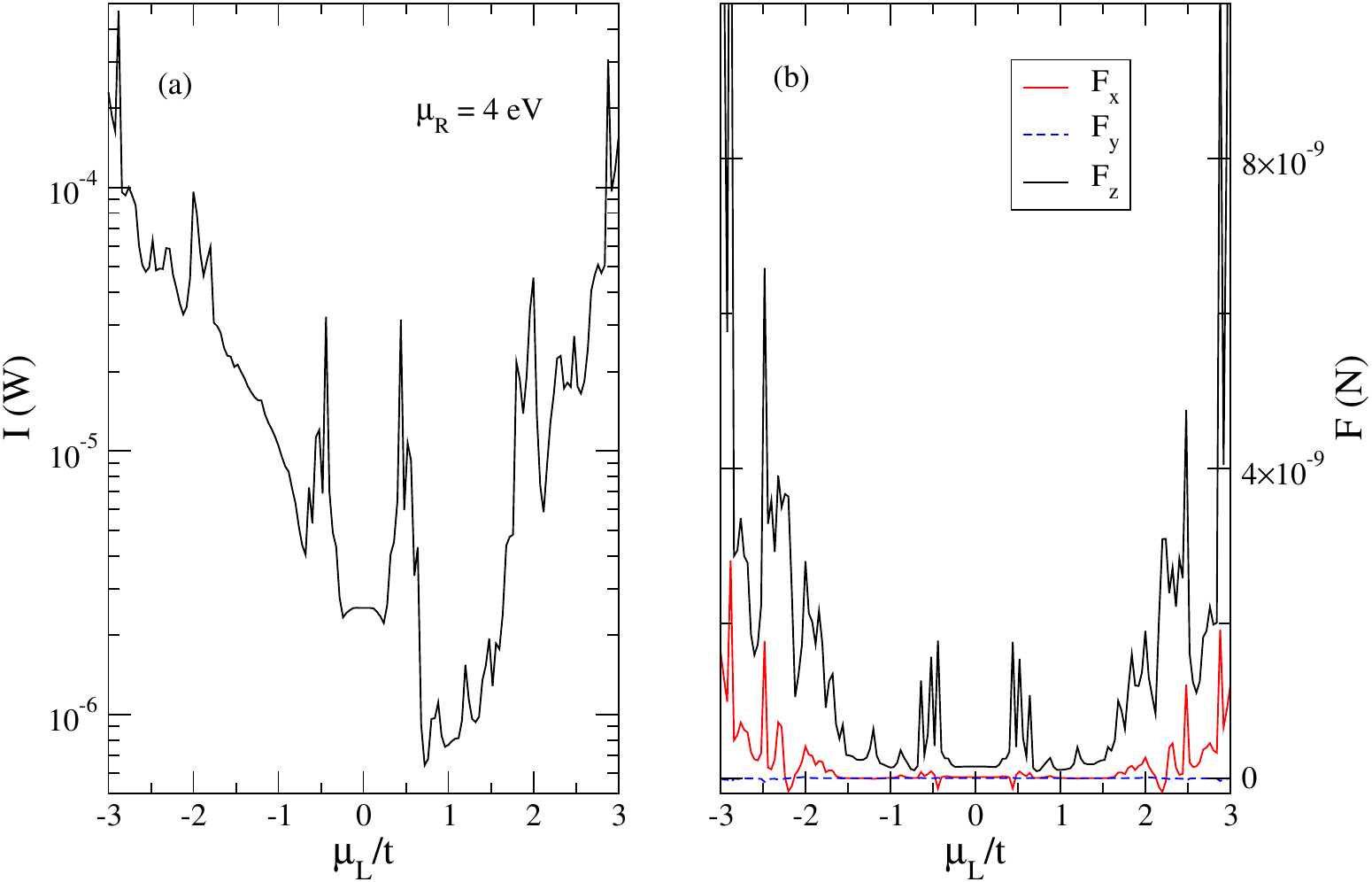}
  \caption{Transport between $13 \times 8$ graphene
strip of armchair edge with C$_{60}$ molecule with a center distance $d=1\,$nm away from the graphene at 300\,K.   The chemical potential of the right side is fixed at
$\mu_R = 4\,$eV, while scanning the left chemical potential $\mu_L$. 
(a) Energy transferred to C$_{60}$,  and (b)  force in $x$ (red), $y$ (dash), and $z$ (solid black) directions to graphene. 
}
 \label{strip-C60-R4eV}
\end{figure}   

Figure~\ref{strip-C60-R4eV} is similar to Fig.~\ref{strip-C60-sym} except now we
fix the right bath at $\mu_R = 4\,$eV and scan the value of $\mu_L$.   This is not symmetric
about 0 or 4\,eV.   The general feature is the same, i.e., driven current causes a large
energy transfer and forces.   The attractive force in $z$ direction is still the largest,
but now there is a small force in $x$ direction,  the sign of which can change.  The
force in the $y$ direction is still negligibly small.   It is clear from the two parameter scans 
of the chemical potentials that metallic systems give large
contributions for both power and force because we have a lot of electrons that can fluctuate, thus inducing
interactions. 

\begin{figure} 
  \centering
  \includegraphics[width=0.9\columnwidth]{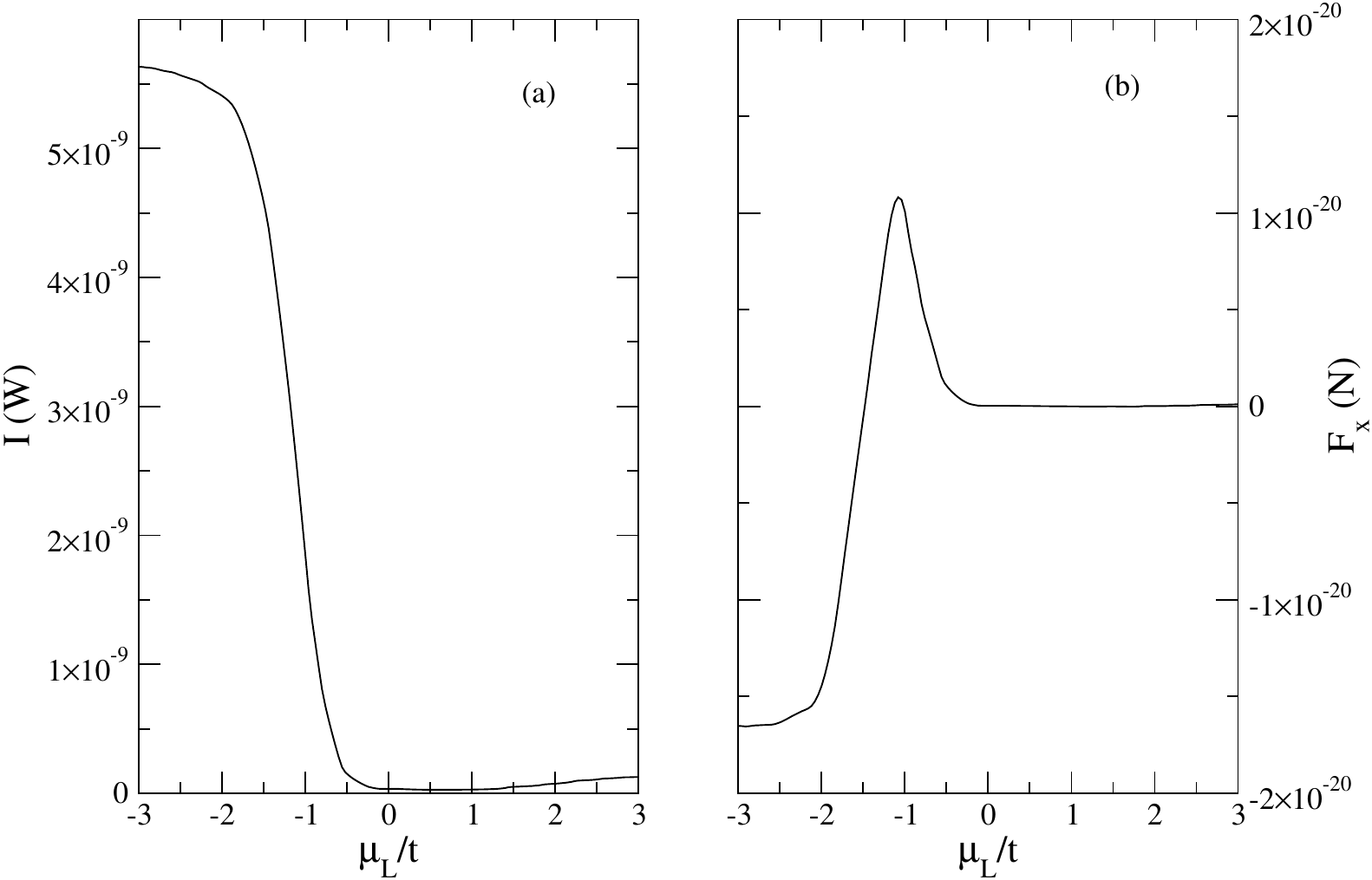}
  \caption{Transport of combined $13 \times 8$ graphene
strip and C$_{60}$ molecule to infinity at 300\,K.   The chemical potential of the right side is fixed at
$\mu_R = 4\,$eV, while scanning the left chemical potential $\mu_L$. 
(a) Energy transferred to infinity,  and (b)  force in $x$ direction acted on the objects. 
}
 \label{strip-C60-R4eVinf}
\end{figure}   

The radiation of the conserved quantities from systems such as benzene molecules \cite{zhang_angular_2020}, 
graphene strips \cite{zhang_microscopic_2022}, or twisted bilayer graphene has been investigated \cite{Yongmei-mw-22}. 
We can calculate the transported quantities to infinity using an approximation
$D \approx v$ and a multiple expansion of the distances for the Green's function.  The
resulting formulas are \cite{zhang_angular_2020,zhang_microscopic_2022,wang_transport_2023},
\begin{eqnarray}
-I_\infty &=& \int_0^\infty\!\! d\omega\, \frac{-\hbar \omega^2}{6\pi^2 \epsilon_0 c^3} {\rm Im\,} \sum_{l,l',\mu}\Pi^<_{l\mu,l'\mu}(\omega),\\
F^\mu_\infty &= & \int_0^\infty\!\!\! d\omega { \hbar \omega^3 \over  60 \pi^2 \epsilon_0 c^5 }  
\sum_{\alpha,l,l'}\Bigg[ 4\,  \Pi^{<}_{l\alpha,l'\alpha}(R^\mu_l -  R^\mu_{l'}) \nonumber\\
&& -  (R^\alpha_l - R^\alpha_{l'}) \Pi^{<}_{l\alpha,l'\mu} 
 -  \Pi^{<}_{l\mu,l'\alpha} (R^\alpha_l - R^\alpha_{l'}) \Bigg].\qquad 
 \label{force-to-inf}
\end{eqnarray}
As expected, from Fig.~\ref{strip-C60-R4eVinf}, we see that the energy and force scales are many orders
of magnitudes smaller than the inter-object transfer.   Due to the structural symmetry, we do not have
angular momentum emission, and the force only has an $x$ component.   Large transfers are  
generated only at a large bias comparable to the hopping parameter $t = 2.7\,$eV.   Note that the
direction of the force changes sign with the bias. 

In our formula, Eq.~(\ref{force-to-inf}), the mutual interaction between C$_{60}$ and graphene is ignored. Under this approximation, each object contributes separately and symmetrically in the $z$ direction, thus, no force in $z$.  In a more precise theory taking the system as a whole and considering the multiple reflections between the objects, it should have a nonzero $z$ component in force due to structural asymmetry.
  
\section{Power, force, and torque between two identical graphene strips}

\begin{figure} 
  \centering
  \includegraphics[width=0.9\columnwidth]{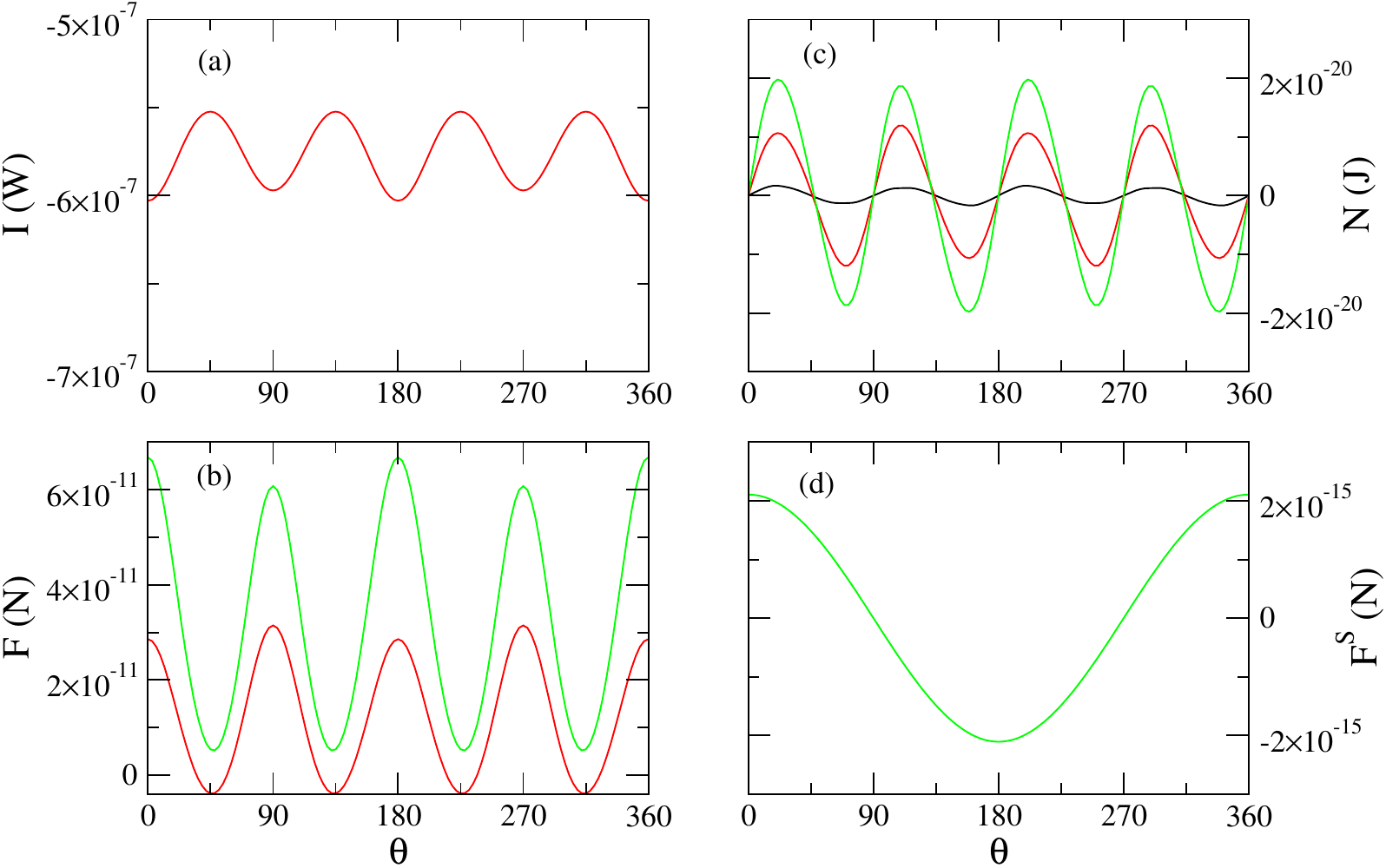}
  \caption{Transfer of the conserved quantities between two structurally identical $13 \times 8$ graphene nanoribbons at 300\,K,
plotted against the rotation angle from 0 to 360 degrees.  For all the subplots, black: $\mu_L = \mu_R = 0$ unbiased 
for both layers; red: unbiased for the bottom layer, symmetrically biased to the top layer  $\mu_L  = -\mu_R = -1\,$eV;
green: both layers are biased by $\mu_L = - \mu_R = -1\,$eV.   The quantities all refer to the bottom layer.  
(a) Energy transferred out of the bottom layer. (b) Fluctuational force. (c) Fluctuational torque. (d) 
Static force based on the Biot-Savart law when both layers are symmetrically biased.}
 \label{fig-two-layer}
\end{figure} 

Finally, in this last section, we present the results of energy, momentum, and angular momentum transfer between
two structurally identical strips, except that the top layer is rotated with respect to the bottom one by an angle $\theta$. 
The two layers are distanced with $d = 1\,$nm.   In Fig.~\ref{fig-two-layer}, we plot the quantities against the
rotation angle.   For the energy transfer, when both strips are in local thermal equilibrium at the same temperature
of 300\,K, there is no heat transfer (numerically, we get values of the order $10^{-24}\,$W, an indication of the numerical
accuracy of our method).  When both layers are
biased, the values are still very small, of the order $10^{-13}\,$W.   The transfer is the largest of the order $10^{-7}\,$W
for the case where one of them is at local equilibrium, and the other is biased, as shown in Fig.~\ref{fig-two-layer}(a).

\begin{figure} 
  \centering
  \includegraphics[width=0.8\columnwidth]{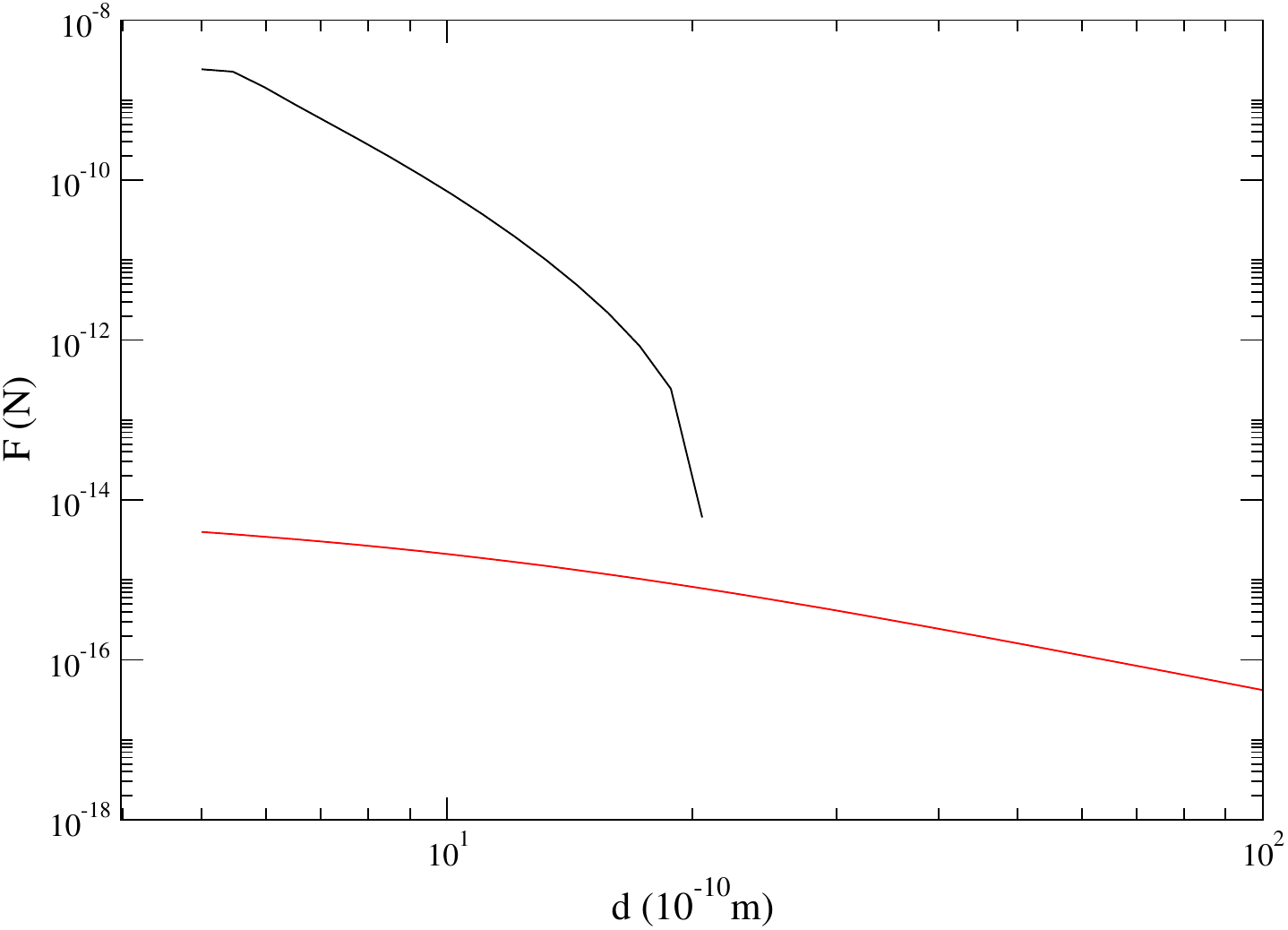}
  \caption{The fluctuational force (black line) vs.~static force (red line) based on the Biot-Savart law, as a function 
of the distance $d$.  The system is two identical $13 \times 8$ graphene strips both at a symmetric bias with 
parallel currents.  The chemical potentials are $\mu_L = - \mu_R = -1\,$eV, 
generating a current of $1.2\times 10^{-4}\,$A.  The temperature is at 300\,K.}
 \label{fig-crossover}
\end{figure} 

In Fig.~{\ref{fig-two-layer}(b) the flucutuational force is plotted.   The forces are the largest when the system is on top of each other at angles 0 or $90^\circ$.   It is the smallest when it is $45^\circ$.  The torque is 0 when it is 
a multiple of $45^\circ$, and oscillating with a period of $90^\circ$.
 Curiously, whether the currents are parallel or anti-parallel, the sign
of the force does not change and remains attractive.  We realized that there is another contribution to the force, which is the explicit effect of
the current by the Biot-Savart law.   This additional correction term is calculated; see Fig.~{\ref{fig-two-layer}(d).  At a distance of 1\,nm, the static force, in 
fact, is 4 orders of magnitude smaller, mainly because of the $1/c^2$ factor in the formula, see Eq.~(\ref{eq-biot-savart}).  

The force between two rotated strips can become repulsive for certain angles.   This anomaly appears
generically also for other small structures, such as between two $2\times 4$ armchair ribbons without the leads. 
We will discuss this issue in the summary section.
    
At a distance $d=1\,$nm between the graphene ribbons, the fluctuational force dominates, but it decays much faster as $d^{-7}$, while the static force decays as
$d^{-2}$.   We can give a rough estimate of the cross-over distance.   We approximate the van der Waals
force by $F_f = E_0 V^2/d^7$, where $E_0$ is an energy scale of order eV, while $V = a L^2$ is roughly
the volume of the graphene (as the polarizability $\alpha$ is proportional to the volume of the system).  
The electric current is about $I = \sigma \Delta \mu/e$.  We use the conductance $\sigma = e^2/h$, $h$ is
the Planck constant.  So
the force by Biot-Savart law is about $F_b = (IL)^2/(d^2 c^2 \epsilon_0)$.   The cross-over distance is 
obtained by equating the two forces, as 
\begin{equation}
d \sim \left(  { E_0 V^2 4\pi\epsilon_0 c^2 \hbar^2 
\over e^2 L^2 \Delta \mu^2} \right)^{1/5}.
\end{equation}   
Using the values $E_0 \sim 1\,$eV, $L \sim 1.5\,$nm, $V \sim 300\,{\rm \AA}^3$, $\Delta \mu\sim 2\,$eV, we find
$d \sim 3\,$nm.  This is comparable to the actual cross-over distance; see Fig.~\ref{fig-crossover}.
   
\section{Summary}

We have given a few applications of the general theory of photon transport between carbon systems. 
They are the van der Waals force between identical C$_{60}$ molecules, or between C$_{60}$ and
graphene flake, or between current-carrying graphene strips. 
Some more technical details are given, such as the diamagnetic self-energy term.   We also relate the
NEGF formalism with the fluctuational electrodynamics results when the systems are in thermal equilibrium.
The main strength of our formalism is the application in nonequilibrium steady states.   In this work,
we focus on the chemical potential drive, but a nonequilibrium setting with two baths at different temperatures is
straightforward. Our numerical
calculations indicate that a current drive generally enhances the transfer of energy and momentum. 
In addition to the attractive force, the drive can produce extra force in the driven direction. 
When both objects have direct currents, the additional static Lorentz force is calculated.   However, it turns out
that this force is about four orders of magnitude smaller at the nanometer scale.  

Our systems are relatively small.   In order to model larger systems, we need to use shortcuts to reduce
the computational complexity.   For example, we can approximate the self-energy $\Pi$ as local, or model
$\Pi$ by a Drude model.   Another possibility is to use periodicity to go into wavevector space when one or both 
systems are infinitely large.    The method can be applied to many situations, e.g., systems that are
magnetic or in a magnetic field.   We can also see drag effects due to current if the electron Green's
functions are the interacting ones with an additional electron-photon self-energy of $GW$ type. 
A periodically driving Floquet system is genuinely a nonequilibrium situation for which our
formalism is applied with minor revision \cite{Gaomin-Wang23}.

Our numerical approach to the transport problems based on nonequilibrium Green's functions does
pose certain difficulties.  First, the Meir-Wingreen formula, Eq.~(\ref{eq-O-meir-wingreen}), 
contains two terms, which are competing and canceling.  It is an intrinsic effect due to the
nonequilibrium setting.  For local equilibrium systems, alternative equivalent forms, such as 
the Matsubara frequency version, offer much better numerical stability.   Thus, it is instructive to
consider perturbative results analytically, focusing on small deviations from the equilibrium result for certain problems,
such as small chemical potential bias.   Second, there is the divergence at zero distance in the
photon Green's functions.  In many situations, this divergent term can be dropped; for example,
the self-force of an object to itself in equilibrium is zero, but it is not clear we can do such manipulation 
for nonequilibrium systems.  Our approach is to give a cut-off to the distance so that zero distance
is forbidden.  This regularizes the divergence.  In fact, this cut-off has a physical meaning, 
it is related to the screening of the electrons \cite{Abajo-2012} and to the electric field inside a body 
\cite{yaghjian_electric_1980}.   Finally, our random phase approximation combined with the
lowest order diamagnetic term may not give correct local gauge invariance.  This breakdown 
of gauge invariance may be the reason we have seen that for certain configurations, we get repulsive forces,
which is not correct.  To overcome this difficulty, we may have to use other gauges such as 
the multi-polar gauge through the Power-Zienau-Woolley transform \cite{Tannoudji89,Schuler21}.  Even though we have
numerical difficulties for some systems, we believe NEGF is still the only offer to treat nonequilibrium systems
correctly from a fundamental point of view beyond fluctuational electrodynamics.

\section*{Acknowledgements}

We thank Svend-Age Biehs for an enlightening discussion regarding the static force.
J.-S. W is supported by an MOE tier 1 grant A-8000990-00-00.  He thanks the group ``Theory of Light-Matter and Quantum Phenomena'' of the Laboratoire Charles Coulomb for hospitality during his stay in Montpellier, where parts of this work were done.  J.-S. W. and M. A. acknowledge CNRS for financial support.  M. A. acknowledges support from the French Agence Nationale pour la Recherche – ANR (``CAT'' project).

\appendix 
\section{Derivation of the Meir-Wingreen formula}
We give a quick and alternative derivation of the Meir-Wingreen formula.  By eliminating the charge due to 
the continuity equation and integrating by parts under the integral sign, we have the energy current, 
force, and torque, as 
\begin{eqnarray}
I &=& {\rm Tr} \langle \dot{\bf A} {\bf j} \rangle, \\
\label{eq-force-Aj}
{\bf F} &= & {\rm Tr} \langle {\bm \nabla } {\bf A} {\bf j} \rangle, \\
{\bf N} &= & {\rm Tr} \bigl\langle ({\bf r} \times {\bm \nabla } + {\bm \epsilon} ) {\bf A} {\bf j} \bigr\rangle,
\end{eqnarray}
here, the notation ${\rm Tr}$ means integration over the volume of a focusing object (out of several well-separated 
objects) and trace of the directions;  ${\bf A} {\bf j}$ is interpreted as a dyadic formed by the vector potential $\bf A$ and current $\bf j$,
$\bm \epsilon$ is a vector tensor, its $i$-th component is the Levi-Civita symbol $\epsilon_{i\mu\nu}$.  The angular brackets denote a  
steady-state average.   We can write the three formulas as a single formula  ${\rm Tr} \bigl\langle \hat{O}/(-i\hbar) {\bf A}{\bf j}\bigr \rangle$ with an appropriate definition of the operator $\hat{O}$.  The energy current is due to the 
Joule heating, the force and torque are consequences of the Lorentz force. 

We can relate the current density to the vector field as ${\bf j} = - v^{-1} {\bf A} = - {\bf A} v^{-1}$.  Here $v^{-1} = -\epsilon_0 ( \partial^2/\partial t^2 + c^2 {\bm \nabla} \times {\bm \nabla} \times \cdot)$ is a differential operator. Since $v$ is symmetric,
acting from the left or right to $\bf A$ is the same.  Using the acting-from-right form, we find
${\rm Tr} \langle \hat{O}  {\bf A} {\bf j} \rangle = -{\rm Tr}\,  \hat{O} \langle {\bf A} {\bf A} \rangle v^{-1}$.
The thermal average $\langle {\bf A} {\bf A} \rangle$ is a short-hand notation for
$\langle A_\mu({\bf r},t) A_{\nu}({\bf r'},t) \rangle$.
The field correlation will be then interpreted as the symmetric version, and can be expressed by 
$\langle {\bf A}{\bf A}\rangle = i\hbar D^K(0)/2$.   Here, the two $\bf A$'s are at different space locations $\bf r$ or
orientations $\mu$ but at the same time, so the Green's function in the time domain is at $t=0$.   By working
in the Fourier space, we obtain $D^K(0)$  by integrating over frequencies.  It is sufficient to integrate over the 
positive frequencies and multiply by 2 due to symmetry in $D^K$.   So we can write the transported quantities as
\begin{equation}
{d\langle \hat{O}\rangle \over dt} = {\rm Re} \int_0^\infty \frac{d\omega}{2\pi} {\rm Tr} \left(  \hat{O} D^K v^{-1} \right).
\end{equation}
We eliminate the operator $v^{-1}$ in favor of Green's functions of $D$ and $\Pi$, by the Dyson equation.
$D^r = v + v \Pi^r D^r$ implies $v^{-1} D^r = I +  \Pi^r D^r$.  Here $I$ is the identity operator in
$({\bf r}, \mu)$, $\omega$ space.
 Taking the Hermitian conjugate, we get
$D^a v^{-1} = I + D^a \Pi^a$.  Using the Keldysh equation $D^K = D^r \Pi^K D^a$, and acting 
by $v^{-1}$ from the right, we find
\begin{equation}
{d\langle \hat{O}\rangle \over dt} =
{\rm Re} \int_0^\infty \frac{d\omega}{2\pi} {\rm Tr} \left[  \hat{O} ( D^r \Pi^K + D^K \Pi^a) \right].
\end{equation} 
Although the self-energy $\Pi$ here is for all objects, we note that the integration is over only the focused object
$\alpha$; as a result, we can replace $\Pi$ by $\Pi_\alpha$ in the above formula.   This is the Meir-Wingreen
formula \cite{Meir92,zhang_microscopic_2022,wang_transport_2023}.

The above derivation is valid when $\langle {\bf A}\rangle = 0$.  When it is not zero, the Green's function $D$ appearing
in the Dyson equation is the centered one or connected one.  We must also take into account a time-independent
static piece,
\begin{equation}
\langle {\bf A} {\bf A} \rangle = i \hbar D + \langle {\bf A}\rangle \langle {\bf A} \rangle.
\end{equation}
This static term does not affect the energy transfer as $\hat{O}$ is a time derivative to the first $\bf A$.  But
it does have a contribution to the force and torque, an explicit static Lorentz force effect. 

To evaluate this static term, we go back to the current; the extra contribution is 
$-\frac{1}{i\hbar} {\rm Tr} \bigl(  \hat{O} \langle {\bf A}\rangle \langle {\bf j} \rangle \bigr)$.
In our discrete representation, the current at site $l$ integrated over a cell volume is 
$-c^\dagger {\bf M}^l c$, where ${\bf M}$ has been defined by Eq.~(\ref{eqMmatrix}) in the main texts by the velocity
matrix.  We use ${\bf A} = - v {\bf j}$.  To the lowest order in the electron-photon interaction, we calculate  
the discrete current as a column vector $\bar I$ with component $(l,\mu)$, 
\begin{equation}
\bar{I}^{l,\mu} = -  2\sum_{j,k} \langle c^\dagger_j M_{jk}^{l\mu} c_k \rangle =2 i\hbar {\rm Tr} \bigl( G^<(0) M^{l\mu} \bigr).
\end{equation}
Then we can compute the extra static term to be
\begin{equation}
\label{eqstaticforce}
\frac{1}{i\hbar} {\rm Tr}\left[ \hat{O} v(\omega\!=\!0) \bar{I} (\bar{I}^\alpha)^T \right].
\end{equation}
Here $\bar I$ is the current of all objects, while ${\bar I}^\alpha$ is the contribution from the focused object $\alpha$. 
Note that if the object $\alpha$ does not carry current, the correction term is 0.  The meaning of trace is changed
to sum over site index and direction, as we are using a discrete current located on each site. 
 
Apparently, Eq.~(\ref{eqstaticforce}) is divergent due to $1/\omega$ and $1/\omega^2$ terms in $v(\omega)$,
see Eq.~(\ref{eq-dv}) in the main texts. 
The origin of this divergence stems from the fact that the continuity equation does not determine the static charge
at $\omega \to 0$.   As a result, the relation $\langle \rho \dot{\bf A} \rangle = - \langle \dot{\rho} {\bf A}\rangle$
breaks down in the static limit.  How much static charge we have has to be a separate model
assumption.  Since the system must be neutral with the explicit charge of the electrons and the ionic background, 
we demand that $\langle \rho \rangle = 0$.  As a result, the electric field term  $\rho {\bf E}$ cannot cancel 
the $-({\bf j} \cdot {\bm \nabla}) {\bf A}$ term leading to Eq.~(\ref{eq-force-Aj}).  When this extra term is 
added, the divergent terms cancel,  and we obtain the Biot-Savart law of a pure magnetic field force,
\begin{equation}
\label{eq-biot-savart}
{\bf F}_\alpha^S = \frac{\mu_0}{4\pi} \sum_{l,l'} {\bar{\bf I}^l_\alpha \times ( \bar{\bf I}^{l'} \times {\hat{\bf R}})
\over R^2 },
\end{equation}
here $\hat {\bf R} = ({\bf R}_l  - {\bf R}_{l'})/R$ is the unit vector from the source (primed quantities) to the observation point.  The torque is similarly calculated by ${\bf R}_l$ cross-product with each term in the sum.

\section{From Meir-Wingreen formula to Casimir-Polder formula}
In this appendix, we present a derivation of the well-known Casimir-Polder formula for the van der Waals coefficient 
$C_6$ from the Meir-Wingreen formula.   Let us consider two objects, calling them $1$ and $2$.  The force
on object $1$ is
\begin{equation}
{\bf F}_1 = {\rm Re} \int_0^\infty \frac{d\omega}{2\pi} {\rm Tr}\Bigl[ \frac{\hbar}{i} {\bf {\bm \nabla}} \left(
D^r \Pi^K_1 + D^{K} \Pi^a_1\right) \Bigr].
\end{equation}
Here the trace means summing over the sites and directions.  To derive the Casimir-Polder formula, we make some
simplifications.  The first step is to solve the retarded Dyson equation in block matrix form:
\begin{eqnarray}
\left(\begin{array}{cc}
D_{11} & D_{12} \\
D_{21} & D_{22} 
\end{array}\right) = 
\left(\begin{array}{cc}
v_{11} & v_{12} \\
v_{21} & v_{22} 
\end{array}\right) + \qquad\qquad\qquad\nonumber \\
\left(\begin{array}{cc}
v_{11} & v_{12} \\
v_{21} & v_{22} 
\end{array}\right)
\left(\begin{array}{cc}
\Pi_{1} & 0 \\
0 & \Pi_{2} 
\end{array}\right)
\left(\begin{array}{cc}
D_{11} & D_{12} \\
D_{21} & D_{22} 
\end{array}\right) .
\end{eqnarray}
Here, the matrix $\Pi$ representing the material properties is bock-diagonal.  The solution can be explicitly found to be
\begin{eqnarray}
D_{11} &=& \epsilon_1^{-1}\bigl(1-v_{12} \chi_2 v_{21} \chi_1\bigr)^{-1}(v_{11} + v_{12} \chi_2 v_{21}), \quad\\
D_{12} &=& \epsilon_1^{-1}\bigl(1-v_{12} \chi_2 v_{21} \chi_1\bigr)^{-1} v_{12} (\epsilon_2^T)^{-1}, 
\end{eqnarray}
where we define the dielectric matrix $\epsilon_1 = 1 - v_{11} \Pi_1$, and the susceptibility $\chi_\alpha= \Pi_\alpha \epsilon_\alpha^{-1}$, $\alpha = 1$, 2.  The superscript $T$ is the matrix transpose (in site and direction space). 
We assume that the distance between the two objects is large, so $v_{12}$ and $v_{21}$ are small, but
$v_{11}$ and $v_{22}$ are not small.  They produce a screening effect. 
In the limit of long distance, small $\chi_\alpha$, but 
$v_{\alpha\alpha} \Pi_\alpha$ finite, it is sufficient
to keep
\begin{eqnarray}
D_{11}  &=& \epsilon_1^{-1} v_{11} + \epsilon_1^{-1} v_{12} \chi_2 v_{21} (\epsilon_1^T)^{-1}  + \cdots,\\
D_{21} &=& \epsilon_2^{-1} v_{21}(\epsilon_1^T)^{-1} + \cdots.
\end{eqnarray}
The derivative of the Green's function is obtained by taking the derivative of the Dyson equation, 
${'D} = {'v} + {'v} \Pi D$, where the prime means partial derivative with respect to space of the first argument. 
To the same order of accuracy, we have, for the derivatives of the photon Green's function,
\begin{eqnarray}
'D_{11}  &=& \Bigl[ {'v}_{11} \bigl(1 + \chi_1 v_{12} \chi_2 v_{21}\bigr) + {'v}_{12} \chi_2 v_{21} \Bigr]  (\epsilon_1^T)^{-1}+ \cdots, \nonumber\\
'D_{12} &=& \bigl( {'v}_{12} + {'v}_{11} \chi_1 v_{12}\bigr)  (\epsilon_2^T)^{-1} + \cdots.
\end{eqnarray}

Putting these expressions into the force formula, we find two types of expressions, these involving only
object $1$; such terms must be zero as a single object has no force on itself.  The terms which are a
product of objects $1$ and $2$ in the self-energies are contributions to the mutual interaction force. 
We take the zero-temperature limit. Then $\Pi_\alpha^K = \Pi_\alpha^{<} + \Pi_\alpha^{>} =  \Pi^r_\alpha - \Pi^a_\alpha  = 2 i \, {\rm Im}\, \Pi^r_\alpha$.   Here, we assume reciprocity in
the sense $(\Pi^r)^T = \Pi^r$.   We can express the force then as
\begin{equation}
{\bf F}_1 = {\rm Im} \int_0^\infty \frac{d\omega}{\pi} 
 {\rm Tr}\Bigl[ \hbar\left( {\bm \nabla} 
v_{12}\right) \chi^r_2 v_{21} \chi^r_1  \Bigr].
\label{eqF1}
\end{equation}
In deriving the above, we have used the identity $\chi (\epsilon^{-1})^* - (\epsilon^T)^{-1} \chi^* = \chi - \chi^*$.
The two extra terms $'v_{11} \chi_1 + {'v_{11}} \chi_1 v_{12} \chi_2 v_{21} \chi_1$ are zero due to symmetry. 
Such terms take the form $'v_{11} F$, here $F^T = F$, $v^T_{11} = v_{11}$, $v_{11}({\bf r}, {\bf r}') = 
v_{11}({\bf r}', {\bf r})$, and $'v_{11} = -v_{11}'$.   Here ${'v}$ denotes derivative to the first argument, and $v'$ is
derivative to the second argument. Using integration by parts (due to the trace sign),
we can move the space derivative around in two ways.  We find $v_{11} F' = - v_{11} F'$, so it must be
zero exactly.   For near distances, if we take into account multiple reflections, the formula (\ref{eqF1}) needs to
be revised by a replacement $v_{21} \mapsto (1-v_{21} \chi_1 v_{12} \chi_2)^{-1} v_{21}$ \cite{Krueger12prb}. 

We take the non-retardation limit (that is, $c\to \infty$).  The free field Green's function simplifies to
\begin{equation}
v = v^r = v^a \approx \frac{1}{4 \pi \epsilon_0 \omega^2 r^3} (\stackrel{\leftrightarrow}{\bf U} - 3 \hat{\bf R}\hat{\bf R}). 
\end{equation}
In this limit, $v$ is real, representing the dipole interaction.  
We assume the distance $r$ between the two objects is much larger than the sizes of the molecules,
so in evaluating $v$, we can take the second molecule all at the origin and the first molecule all at $(0,0,r)$. 
As a result, $v_{jk}$ no longer depends on the site indices.   The summation over the sites is carried out 
only for the self-energies.  We also assume that the site summed $\chi^r_\alpha$ is isotropic and is proportional
to the identity.   We are able to relate the ``total'' $\Pi^r \epsilon^{-1}$ to the isotropic polarizability $\bar{\alpha}$.  
The site summed $\chi^r$ is the linear response to the volume integrated current density, $\int j_\mu dV = 
- \sum_{l,l',\nu} \chi^r_{l\mu,l'\nu}A^{\rm ext}_\nu$.   We can write the current as charge density times velocity,
which  in turn can be written as the rate of change of dipole moment; thus, $(-i\omega){\bf p} = - \frac{1}{3}
\chi^r(\omega)_{\rm tot} {\bf A}$.   Here, we define a scalar quantity
\begin{equation}
\chi^r_{\rm tot}(\omega) =  \sum_{l,l',\mu} \chi_{l\mu,l'\mu}^r(\omega) = -3 \omega^2 \bar{\alpha}(\omega).
\end{equation}
The $\omega^2$ factor is due to $\frac{d}{dt}{\bf p} =  \int {\bf j}\, dV$, ${\bf E} = - d{\bf A}/dt$, and
in the frequency domain, the time derivative is $-i\omega$.   Replacing $\chi^r$ by the isotropic $\bar{\alpha}$,
multiplying the matrices in the $x,y,z$ direction space, and taking the trace (which gives a factor of $6$), we obtain,
\begin{equation}
\label{C6realf}
F^z_1 = \frac{3\hbar}{\pi} \int_0^{\infty}\!\! d\omega \left(\frac{1}{4 \pi \epsilon_0}\right)^2 
{\rm Im} \Bigl[ \bar{\alpha}_1(\omega) \bar{\alpha}_2 (\omega) \Bigr]  \left({-6 \over r^7}\right).
\end{equation}
We can now identify the coefficient $C_6$ as the factor in front of $\left(-6/r^7\right)$.  The final step is to make
a Wick's rotation by integrating over the positive imaginary axis from $0$ to $i \infty$.  Since the retarded
response function or product of retarded Green's functions are still retarded, the integrand is analytic 
on the upper half-plane.   This leads to the final Casimir-Polder expression in the non-retarded limit as \cite{Casimir_Polder_1948},
 \begin{equation}
C_6 = \frac{3\hbar}{\pi}\left(\frac{1}{4 \pi \epsilon_0}\right)^2  \int_0^{\infty}\!\! d\omega' 
 \bar{\alpha}_1(i\omega') \bar{\alpha}_2 (i\omega')  .
\end{equation}
In doing this, taking the imaginary part becomes taking the real part, as $\omega = i \omega'$, but
the analytically continued response functions are real on the imaginary axis. 

We have used Eq.~(\ref{C6realf}) of the real frequency formula to compute $C_6$ for the C$_{60}$ with
C$_{60}$ interaction.  The numerical value is $\sim 3.07\times 10^{4}$ a.u., in good
agreement with a direct fit to the distance dependence of the force.

\clearpage
\bibliography{fuller-abbre-withtitle}

\end{document}